\documentclass[twocolumn,prl,floatfix,preprintnumbers,a4paper,nofootinbib,superscriptaddress]{revtex4-1}

\usepackage{epsfig}
\usepackage{graphics}
\usepackage{graphicx}
\usepackage{amsmath,amssymb}
\usepackage{amsfonts}
\usepackage{bm}
\usepackage[usenames,dvipsnames]{color}
\usepackage{amssymb}
\usepackage{mathtools}
\usepackage{parskip}
\usepackage{psfrag}
\usepackage{times}
\usepackage[varg]{txfonts}
\usepackage[colorlinks, pdfborder={0 0 0}]{hyperref}
\definecolor{LinkColor}{rgb}{0.75, 0, 0}
\definecolor{CiteColor}{rgb}{0, 0.5, 0.5}
\definecolor{UrlColor}{rgb}{0, 0, 0.75}
\hypersetup{linkcolor=LinkColor}
\hypersetup{citecolor=CiteColor}
\hypersetup{urlcolor=UrlColor}
\usepackage[utf8]{inputenc}
\usepackage{ulem}
\usepackage{hyperref}
\usepackage{float}
\usepackage{soul}
\usepackage[hyphenbreaks]{breakurl}

\def\be{\begin{equation}}
\def\ee{\end{equation}}
\def\bea{\begin{eqnarray}}
\def\eea{\end{eqnarray}}

\newcommand\apjs{Astrophys. J. Suppl. Ser.}
\newcommand\apjl{Astrophys. J. Lett.}
\newcommand\mnras{ Mon. Not. R. Astron. Soc.}

\begin{document}

\title{Accretion-induced spin-wandering effects on the neutron star in Scorpius X-1: 
Implications for continuous gravitational wave searches}

\author{Arunava Mukherjee}
\email[Email: ]{arunava.mukherjee@aei.mpg.de}
\affiliation{Max-Planck-Institut f{\"u}r Gravitationsphysik, Callinstr. 38, D-30167 Hannover, Germany}
\affiliation{International Centre for Theoretical Sciences, Tata Institute of Fundamental Research, Bangalore 560012, India}
\author{Chris Messenger}
\email[Email: ]{Christopher.Messenger@glasgow.ac.uk}
\affiliation{SUPA, School of Physics and Astronomy, University of Glasgow, Glasgow G12 8QQ, United Kingdom}
\author{Keith Riles}
\email[Email: ]{kriles@umich.edu}
\affiliation{University of Michigan, Ann Arbor, MI 48109, USA}

%
%
%
%


\begin{abstract}
The LIGO's discovery of binary black hole mergers has opened up a new era 
of transient gravitational wave astronomy. The potential detection of
gravitational radiation from another class of astronomical objects, rapidly
spinning non-axisymmetric neutron stars, would constitute a new area of
gravitational wave astronomy. Scorpius X-1 (Sco X-1) is one of the most
promising sources of continuous gravitational radiation to be detected with
present-generation ground-based gravitational wave detectors, such as Advanced
LIGO and Advanced Virgo. As the sensitivity of these detectors improve in the
coming years, so will power of the search algorithms being used to find
gravitational wave signals. Those searches will still require integation over
nearly year long observational spans to detect the incredibly weak signals 
from rotating neutron stars. For low mass X-ray binaries such as Sco X-1 this
difficult task is compounded by neutron star ``spin wandering" caused by
stochastic accretion fluctuations. In this paper, we analyze X-ray data from
the {\it RXTE} satellite to infer the fluctuating torque on the neutron star 
in Sco X-1. We then perform a large-scale simulation to quantify the statistical
properties of spin-wandering effects on the gravitational wave signal frequency
and phase evolution. We find that there are a broad range of expected maximum
levels of frequency wandering corresponding to maximum drifts of between
$0.3$--$50$ $\mu$Hz/sec over a year at 99\% confidence. These results can be
cast in terms of the maximum allowed length of a coherent signal model
neglecting spin-wandering effects as ranging between $5$--$80$ days. This study
is designed to guide the development and evaluation of Sco X-1 search
algorithms.
\end{abstract}
\maketitle 

\preprint{}
\date{\today}

\section{Introduction and motivation} 
\label{sec:level1}

%
It has long been proposed that rapidly spinning neutron stars could emit
detectable gravitational waves (GWs)~\citep{1998ApJ...501L..89B} and that neutron
stars in low-mass X-ray binary (LMXB) systems form an especially interesting
class of objects.  The neutron stars in these sytems accrete matter from
their companion stars (\textquotedblleft donor\textquotedblright) and thus 
can be spun-up to spin frequencies $>$ 500 Hz as the accreted matter also carries 
angular momentum to the neutron star \citep{1997ApJS..113..367B}, the amount of 
which is expected to be larger for larger mass accretion rates ($\dot{\text{M}}$). 
Observations show, however, that the fastest spinning accreting neutron star 
has a spin frequency much smaller than the expected maximum allowed value 
(breakup frequency)~\citep{2008AIPC.1068...67C}. GWs have been proposed as a 
braking mechanism that prevents these accreting neutron stars from spinning up 
to higher frequencies~\citep{2017arXiv170507669P, 2008AIPC.1068...67C}.

%
Recent detections of GWs from 4 confirmed binary black hole merger events have
started a new era of transient GW astronomy~\citep{Abbott:2016blz, Abbott:2016nmj, 
2016PhRvX...6d1015A, Abbott:2017vtc, 2017arXiv170909660T}. In the contrasting 
continuous emission scenario, GWs generated from the neutron star within the 
Scorpius X-1 (Sco~X-1) system are an exciting prospect for detection. Sco~X-1 is 
the brightest persistent extra-solar X-ray source in the sky. The high X-ray 
luminosity of Sco~X-1 suggests an accretion rate close to the Eddington 
limit~\citep{1989ESASP.296..215L}. 
Moreover, the distance to the source is also quite small $\approx 2.8$ kpc
~\citep{1999ApJ...512L.121B}. Thus, it was identified early as a potential GW 
target~\citep{1984ApJ...278..345W}. In light of the recently upgraded 
second-generation GW detectors, Advanced LIGO and Advanced VIRGO, it should soon 
be possible to beat the \textquotedblleft torque-balance\textquotedblright\ limit 
over a wide frequency range \citep{2015PhRvD..91j2003L, 2015PhRvD..91j2005W, 
2015PhRvD..92b3006M}. According to this scenario, the accretion-induced spin-up 
torque is primarily balanced by the total spin-down torque due to gravitational 
and electromagnetic (EM) radiation, although other braking mechanisms may also 
play a role \citep{2017arXiv170507669P}.

%
The GW signals emitted from these rapidly spinning neutron stars will be
extremely weak relative to the signals recently detected from binary black
hole mergers~\cite{2016PhRvX...6d1015A}, despite the relative proximity of
galactic neutron star systems. The expected continuous GW strain-signal from 
a spinning neutron star in Sco~X-1 will be of the order of $h \sim 10^{-25}$ 
or smaller~\cite{2015PhRvD..91f2008A}. Detection will likely require an
observation duration of several months to years, with signal-to-noise ratio
(SNR) accumulated via long term integration of the data.

%
One crucial practical problem in the search for GWs from LMXBs is spin wandering 
(SW), which can potentially degrade the effectiveness of continuous GW searches 
from accreting neutron stars. Although the spin-down torque due to GW and EM 
emission is generally unlikely to change by a significant amount during
the observation period ($\sim 1$ year), the mass accretion rate can change
appreciably over this timescale, leading to an appreciable change in the
instantaneous stellar spin-frequency and affect cumulative rotational phase.
Such fluctuations make it challenging to integrate a GW signal coherently and
achieve ideal detection efficiency. In this letter, we estimate for the first
time detailed accretion fluctuation effects on the stellar spin frequency and on
the cumulative rotational phase of the neutron star in the Sco~X-1 system. The
results can be applied to any other accreting neutron stars by choosing the 
values of appropriate physical parameters along with observational data from 
those systems. 

%
In the following sections of this paper, we briefly describe our simulations to
model astrophysically realistic spin-wandering effects on the continuous GW
signals from the neutron star in Sco~X-1. Numerical results are presented for
expected fluctuation effects on the GW frequency ($f_{\text{GW}}$) and GW phase 
($\phi_{\text{GW}}$) as a function of search integration time ranging over timescales 
of $\sim$ hour to greater than a year. 

\section{Fluctuating Torque On The Neutron Star And The Spin Wandering Effect} 
\label{sec:level2}

%
Neutron stars in LMXB systems typically accrete matter from the companion star,
often termed the ``donor", through Roche-lobe overflow and are spun-up by
the matter, a process known as ``recycling'' of neutron stars. According to the 
\citet{1997ApJS..113..367B} model (see Eqn.~7), accretion-induced torque ($N$) 
resulting changes in spin-frequency is given by,
%
\begin{equation}
\begin{split}
\dot{f}_{\text{spin}} = \bigg(\frac{N}{2 \pi I_{\text{spin}}}\bigg)
\simeq 1.6 \times 10^{-13} \bigg(\frac{\dot{M}}{10^{-10} M_{\odot} yr^{-1}}\bigg) 
\times \bigg(\frac{P_{\text{spin}}}{s}\bigg)^{1/3} \\
\times \bigg(\frac{r_{m}}{r_{\text{co}}}\bigg)^{1/2} 
\times \bigg(\frac{I_{0}}{I_{\text{spin}}}\bigg) s^{-2},
\label{master-eqn}
\end{split}
\end{equation}
%
where $I_{\text{spin}}$ is the moment of inertia of the neutron star about the
spin-axis, $I_{0} = 10^{45}$ g-cm$^{2}$, $\dot{M}$ is the mass accretion rate, 
$P_{\text{spin}} = f_{\text{spin}}^{-1}$ is the spin-period, $r_{m}$ is the 
magnetic radius and $r_{\text{co}}$ is the co-rotation radius of the accreting 
neutron star. 

\begin{figure*}
\begin{center}
\begin{tabular}{lr}
\includegraphics[width=0.5\textwidth,angle=0]{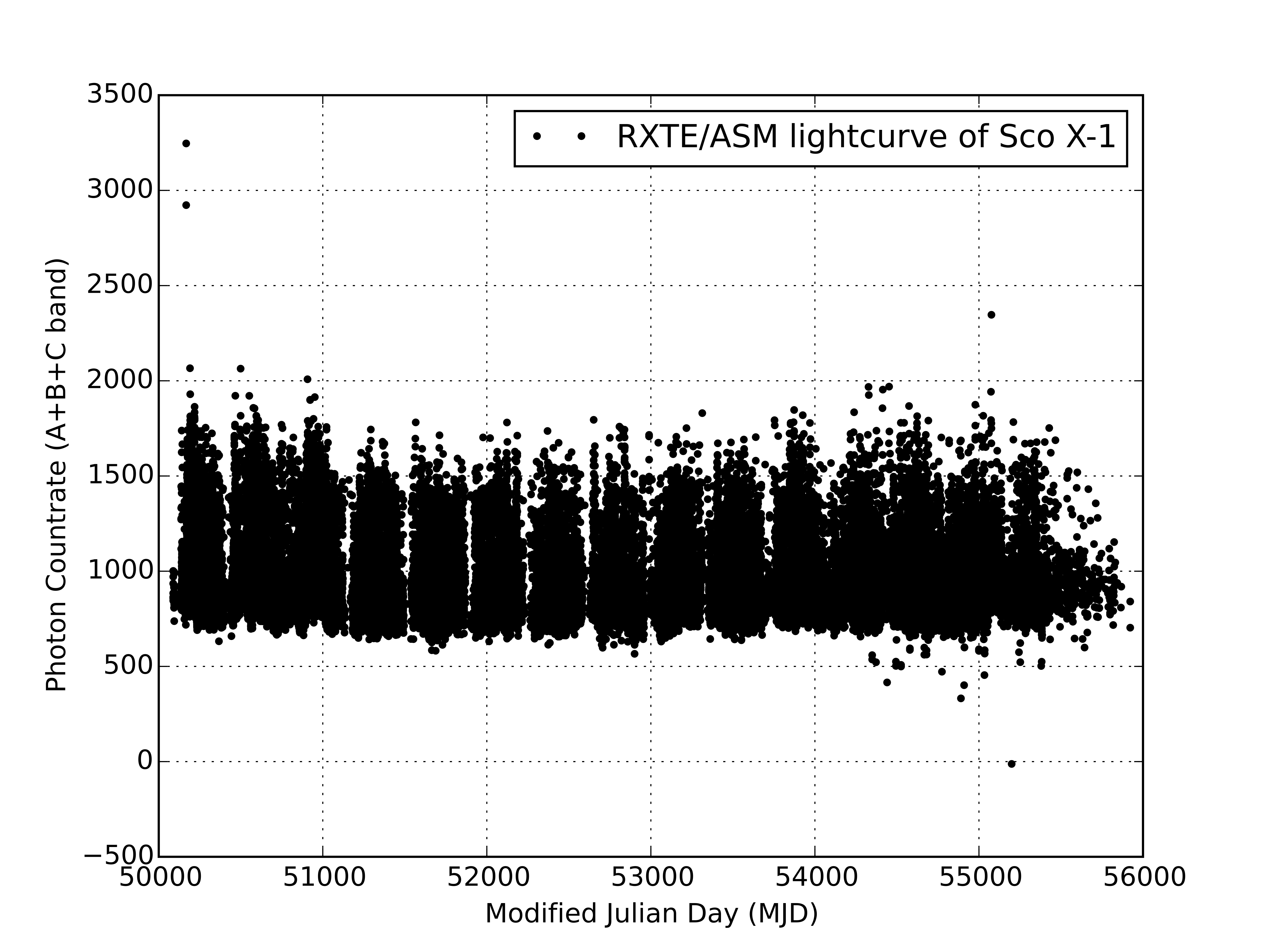} &
\includegraphics[width=0.5\textwidth,angle=0]{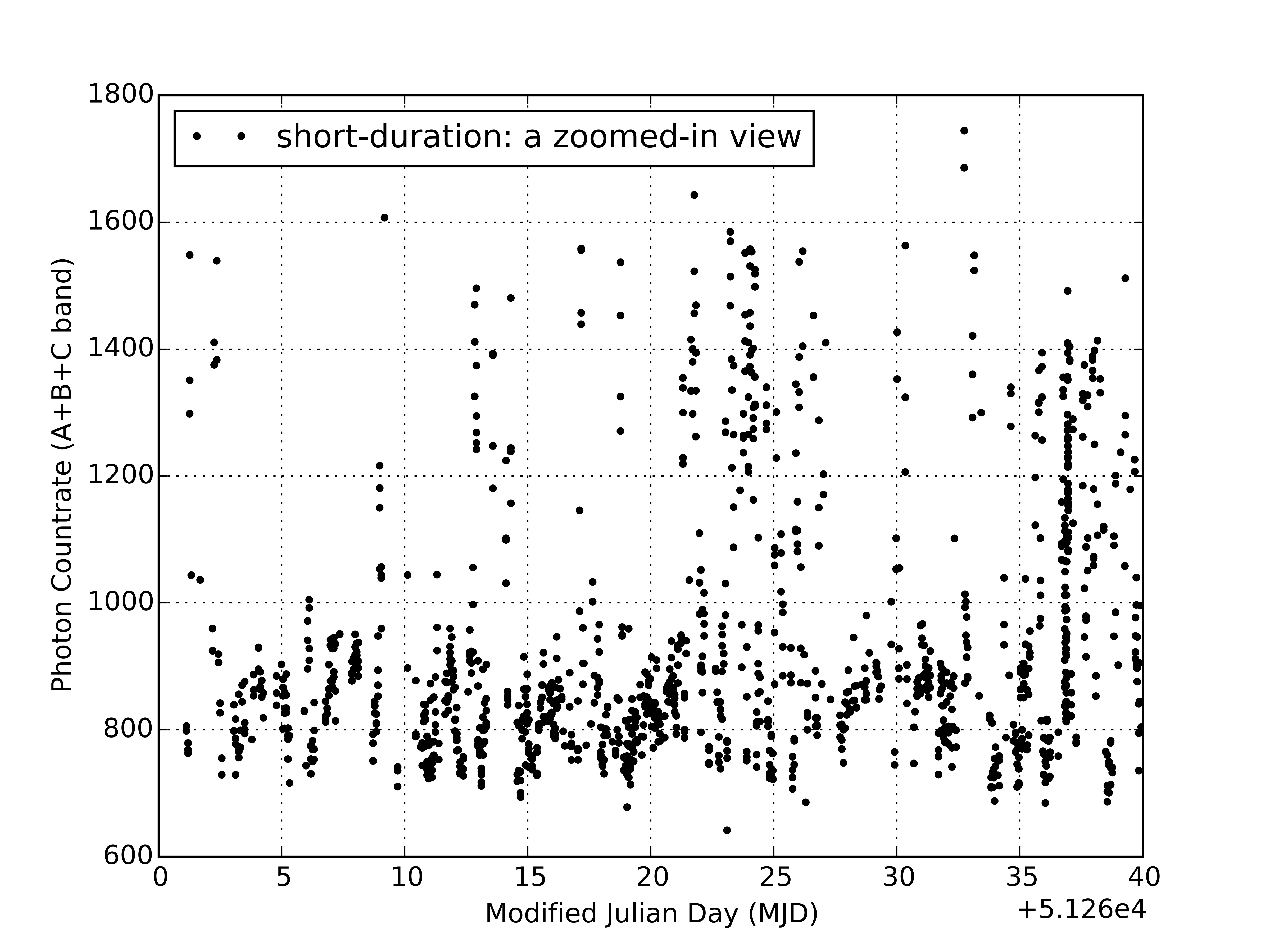}
\end{tabular}
\end{center}
\caption{{\it Left panel:} Sum-band (i.e., A+B+C bands) photon count-rates over nearly 
15 years of {\it RXTE}/ASM monitoring of Sco~X-1. The ASM sum-band covers the photon 
energy range of 2-10 keV. {\it Right panel:} Magnification of one sample 40-day period 
in the left panel.}
\label{ASM-lc}
\end{figure*}

%
According to the torque balance scenario, accretion induced spin-up torque 
on the neutron star is balanced by the spin-down torques from EM and 
gravitational radiation~\citep{2008AIPC.1068...67C, 1998ApJ...501L..89B}. The
spin-down torque due to EM radiation depends mainly on the magnetic dipole 
field of the neutron star, and that due to gravitational radiation depends on 
the quadrupolar deformation of the star. Neither the external magnetic field 
nor the quadrupolar deformation, ignoring time varying excitations such as 
r-mode~\citep{2001IJMPD..10..381A, 2000PhRvD..62h4030L}, 
is likely to change significantly over our longest considered timescale of 
$\sim 1$ yr. Hence we neglect changes in spin-down torques in the following 
analysis.

%
The mass accretion rate ($\dot{M}$) for a neutron star in a typical LMXB system
(including Sco~X-1) can change appreciably on much shorter timescales than
$\sim 1$ yr. The instantaneous mass accretion rate $\dot{M}(t)$ can thus be
split into two parts:
\begin{equation} 
\dot{M}(t)  = \langle \dot{M}(t) \rangle + \Delta \dot{M}(t).
\label{fluc-mean_Mdot} 
\end{equation}

However, $\dot{M}(t)$ is related to the amount of instantaneous torque and 
thus to the spin frequency derivative $\dot{f}(t)$ of the neutron star 
according to Eq.~\ref{fluc-mean_Mdot}, that derivative can be decomposed:
\begin{equation} 
\dot{f}(t)  = \langle \dot{f}(t) \rangle + \Delta \dot{f}(t).
\label{fluc-mean_fdot} 
\end{equation}

The torque balance scenario asserts that, over a long timescale, the spin-up and 
spin-down torques balance each other, i.e., $\langle \dot{f}(t) \rangle = 0$; but 
$\Delta \dot{f}(t) \neq 0$ over a timescale of hours to a few years, because of 
the observed $\Delta \dot{M}(t) \neq 0$ (see Figure~\ref{ASM-lc}).
%

%
As a consequence, the neutron star is out of spin-equilibrium instantaneously
and either spinning up or spinning down as the resultant torque fluctuates. We
call the corresponding fluctuation in spin-frequency relative to the long term
average, the ``spin-wandering" effect.  In this paper, we perform a first
attempt to estimate quantitatively this spin-wandering effect for the neutron
star in Sco~X-1 over relevant GW search timescales and using an 
astrophysically-motivated model together with relevant observational data. 

%
For the purpose of estimating the spin-wandering effect, we therefore require
information regarding the time variation of the mass accretion rate ($\dot{M}(t)$), 
the spin period ($P_{\text{spin}}$), and the magnetic and co-rotation radii 
($r_{m}$ and $r_{\text{co}}$) of the neutron star. Below, we discuss estimates
for these quantities.

{\subsection{\label{subsec2.1} Estimating the mass accretion rate: analyzing X-ray data}

%
Estimating the time-dependent mass accretion rate $\dot{M}(t)$ is an important 
part of this analysis. The X-ray flux from an accreting LMXB is taken as an 
estimator of instantaneous $\dot{M}$, provided we know the distance to the 
source~\cite{1995xrbi.nasa.....L}. The distance to Sco~X-1 has been reported 
to be $2.8 \pm 0.3$ kpc \cite{1999ApJ...512L.121B}, and data from the long 
term monitoring of Sco~X-1 can be utilized to estimate the character of the 
fluctuations $\Delta\dot{M}(t)$. For this purpose we used archival X-ray data 
from the all-sky monitor (ASM) instrument of {\it Rossi X-ray Timing Explorer} 
({\it RXTE}){\footnote {for details see 
\url{https://heasarc.gsfc.nasa.gov/docs/xte/xte_1st.html}}}. For the purpose 
of estimating the time-averaged mass accretion rate $\langle \dot{M}(t) \rangle$, 
and the time dependent variability $\Delta \dot{M}(t)$ around this mean value, 
we performed the following steps. 

%
We analyzed the timeseries of the sum-band (i.e., A+B+C bands) photon
count-rates of Sco~X-1 from the {\it RXTE}/ASM instrument giving us a crude
estimate of $\langle\dot{M}(t)\rangle$ and its fluctuation over a period of
nearly 15 years. The resultant {\it RXTE}/ASM lightcurve is shown in
Fig.~\ref{ASM-lc}.  From the two panels of the figure, one can see that the
mass-accretion rate fluctuates quite dramatically over a wide range of
timescales. The scale of fluctuation in luminosity and correspondingly in the
inferred $\dot{M}$ often become comparable to their respective mean values,
suggesting that fluctuations in accretion induced torque, and consequently the
spin-wandering effect, may be significant. 

%
This sum-band photon count-rate of {\it RXTE}/ASM alone, however, cannot be
used to infer the $\dot{M}(t)$, as (i) the does not have a good energy 
resolution and thus cannot provide X-ray flux accurately, and (ii) it detects 
only photons in the 2-10 keV range. For this purpose, we perform the following 
steps. Since, Sco X-1 is a highly variable source, we cannot compare the flux 
across different bands observed at different epochs. Therefore, we compute the 
95\% confidence level upper-limit and lower limit of the RXTE/ASM count-rate 
to be 1716 photons/s and 618 photons/s, respectively. Comparing the reported 
upper-limit of mass accretion rate from Sco X-1 by \citet{2005ApJ...623.1070M}, 
we find the upper-limit (95\% confidence level) of the flux corresponds to 
$\simeq 0.42 \times 10^{-8} M_{\odot} yr^{-1}$ and lower-limit (95\% confidence 
level) of the flux corresponds to $\simeq 1.2 \times 10^{-8} M_{\odot} yr^{-1}$. 
This corresponds to a conversion factor of $7.03 \times 10^{-10}$ M$_{\odot} yr^{-1}$ 
per 100 photon count-rate in RXTE/ASM.

In parallel, we make an attempt to estimate the X-ray flux from Sco X-1. In order 
to estimate it, we need to analyze spectral data from an instrument with good 
photon energy resolution. Since, the spectral resolution of the ASM instrument 
is limited, we use another instrument, the Proportional Counter Array (PCA) 
{\footnote {for details see \url{https://heasarc.gsfc.nasa.gov/docs/xte/PCA.html}}} 
from the same satellite as a calibration reference. The PCA enables us to find 
a number of (nearly) simultaneous measurements of Sco~X-1 and allows us to estimate 
the energy flux in 2-10 keV from the measured X-ray spectra.

%
We analyzed a few pointed {\it RXTE}/PCA observations that are (near) simultaneous 
to some of the ASM observations of Sco~X-1. We performed the data extraction using 
the standard {\tt Ftools}{\footnote {for details see 
\url{https://heasarc.gsfc.nasa.gov/ftools/ftools_menu.html}}} analysis software 
and background subtraction using {\tt runpcabackest}{\footnote {for details see
\url{https://heasarc.gsfc.nasa.gov/docs/xte/recipes/pcabackest.html}}} tool in
the {\tt Heasoft}{\footnote {for details see
\url{https://heasarc.gsfc.nasa.gov/lheasoft/}}} data analysis software. In each
case, the continuum spectral energy density (SED) in the $2-10$ keV photon
energy band can be described well with a combination of the 
{\tt phabs*(diskbb+compTT+Gaussian)} model{\footnote {for details of each 
individual model see
\url{https://heasarc.gsfc.nasa.gov/xanadu/xspec/manual/XspecModels.html}}}
using the {\tt xspec} spectral analysis package{\footnote {for details of the
spectral analysis package see
\url{https://heasarc.gsfc.nasa.gov/xanadu/xspec/manual/manual.html}}} to obtain
satisfactory ($\chi^{2}/dof \approx 1$){\footnote {dof stands for number of 
degrees of freedom} statistical significance (see Fig.~\ref{PCA-sed}).
From the best-fit models, we have estimated the fluxes in the analyzed $2-10$
keV bands. The observed fluxes scale well (to within $\sim 10\%$) with the
sum-band photon count-rates from simultaneous {\it RXTE}/ASM observations. In
Fig.~\ref{PCA-sed} we show a representative plot with the best-fit theoretical
model along with the corresponding {\it RXTE}/PCA observations. From 
Fig.~\ref{PCA-sed}, we see that most of the X-ray flux of Sco~X-1 comes in this 
$2-10$ keV photon energy range. Thus, it can be taken as a good tracer of 
$\dot{M}(t)$. 

\begin{figure}[h]
\begin{center}
\begin{tabular}{c}
\includegraphics[width=0.5\textwidth,angle=0]{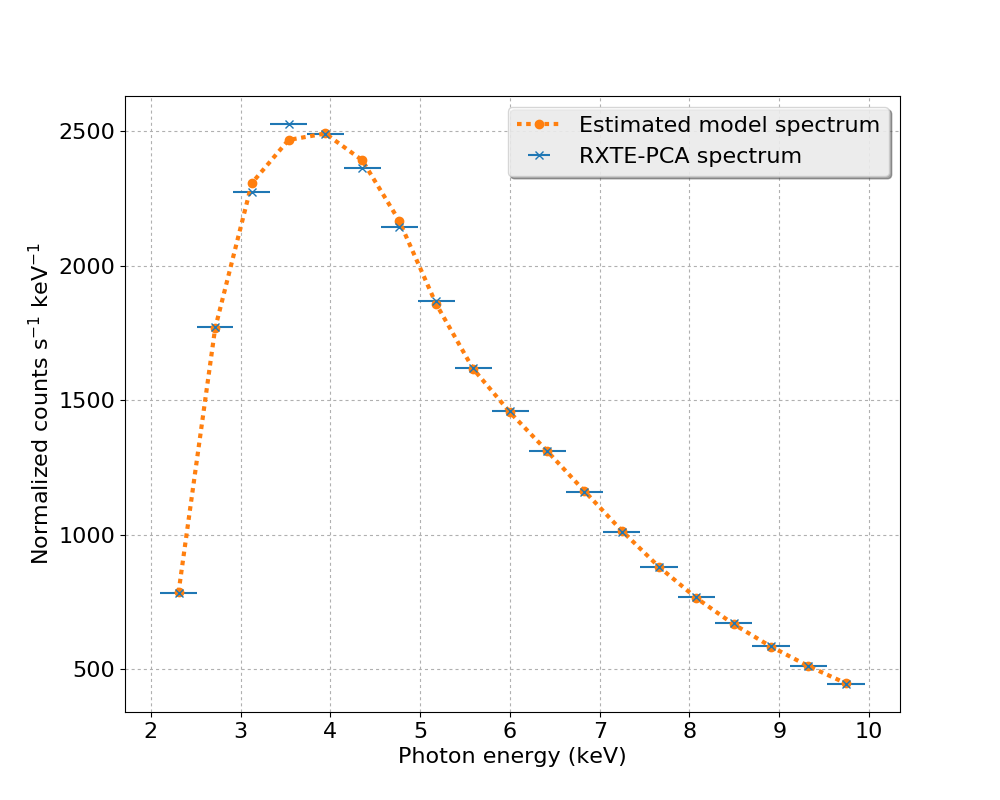}
\end{tabular}
\end{center}
\caption{The spectral energy distribution of Sco~X-1 in the photon energy band of 
2-10 keV observed with the {\it RXTE}/PCA instrument. The observed spectrum is well 
described by a combination of {\tt phabs*(diskbb+compTT+Gaussian)}. The discrete 
(light-blue colored) lines represent data from {\it RXTE}/PCA and the dotted 
continuous (oranged colored) lines are the estimated model spectrum. We have 
performed this model estimation and corresponding X-ray flux using the {\tt xspec} 
spectral analysis package.}
\label{PCA-sed}
\end{figure}

%
However, the estimated flux from this photon energy spectrum corresponds to the
emission only in the $2-10$ keV band, whereas we need to measure the bolometric
flux from Sco~X-1 to estimate the actual $\dot{M}$. In order to achieve this,
we estimate the theoretical flux over a sufficiently large range, $0.1-50.0$
keV, over which most of the radiation from Sco~X-1 is captured, and then use it
to scale the time-series of {\it RXTE}/ASM count-rates. The estimated flux in
the $0.1-50.0$ keV energy band from these observations is found to be $\simeq
3.56 \times 10^{-7}$ erg cm$^{-2}$ s$^{-1}$, which corresponds to $5.9 \times
10^{-9}$ M$_{\odot}$/yr for a source distance of 2.8 kpc. The corresponding
(near) simultaneous ASM photon count-rate corresponds to 956 s$^{-1}$. This 
estimation of $\dot{M}(t)$ of Sco~X-1 from {\it RXTE}/PCA is in good agreement with 
(only $14\%$ below) the corresponding estimation from the ASM countrate calibrated 
to \citet{2005ApJ...623.1070M}. This underestimate in $\dot{M}(t)$ could, possibly, 
due to presence of some other emission mechanisms predominant outside this $2-10$ 
keV band over which we modeled photon energy spectrum which can result in a small 
fraction of the flux falling outside this measured energy band. It is worth 
mentioning that, the uncertainty in distance to Sco~X-1 is $\sim 10\%$ implying 
an uncertainty in absolute flux of $\sim 20\%$. For our purpose, this uncertainty 
in $\dot{M}(t)$ is rather small as compared to other uncertainties in our modelling 
as discussed in section~\ref{sec:level4}.
}

\begin{figure}[h]
\begin{center}
\begin{tabular}{c}
\hspace*{-0.5cm}
\includegraphics[width=0.55\textwidth,angle=0]{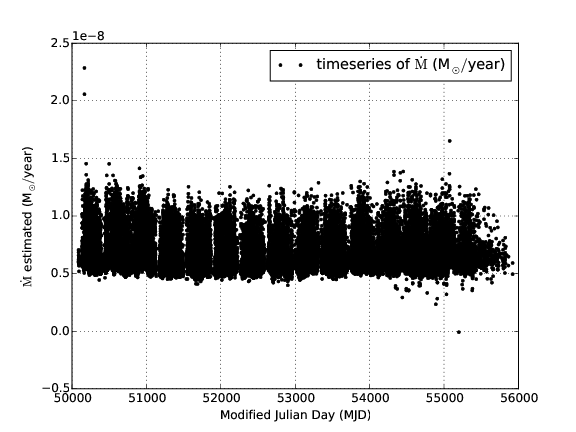}
\end{tabular}
\end{center}
\vspace*{-0.5cm}
\caption{The inferred timeseries of the mass accretion rate $\dot{M}(t)$ from 
Sco~X-1 over approximately 15 years period in Modified Julian Day (MJD).}
\label{ASM-Mdot}
\end{figure}

\subsection{Estimating magnetic radius ($r_{m}$) and co-rotation radius ($r_{\text{co}}$)}
\label{subsec2.2}

%
The magnetic radius ($r_{m}$) and co-rotation radius ($r_{\text{co}}$) play crucial 
roles too, in determining the spin-wandering effect for an accreting neutron star 
in an LMXB system (see Eq.~\ref{master-eqn}). The ratio of these two characteristic 
radii, $(r_{m}/r_{\text{co}})$, determines the coupling between the magnetosphere of 
the neutron star and the accretion disk around it. The magnetic radius length-scale
corresponds to $r_{m} = \xi r_{\text{A}}$; where $\xi$ is expected to be in the 
range $\simeq 0.5-1$. It is not completely certain how much the accreted matter 
is guided by the field lines to fall on to the polar caps. In this paper we take a 
conservative approach, leading to the maximum possible spin-wandering effect. 
Here, $r_{\text{A}}$ is the Alfv\'en radius, quantified in this case as: 

\begin{equation}
\begin{split}
r_{\text{A}} = \bigg(\frac{\mu^{4}}{2GM\dot{M}^{2}}\bigg)^{1/7} \simeq 6.8 \times 10^{8}  \bigg(\frac{\mu}{30 G cm^{3}}\bigg)^{4/7} \\
\times \bigg(\frac{10^{-10} M_{\odot} yr^{-1}}{\dot{M}}\bigg)^{2/7} \times \bigg(\frac{1.4M_{\odot}}{M}\bigg)^{1/7} cm,
\end{split}
\end{equation}
where $\mu$ is the star's magnetic dipole moment and $M$ is the mass of the 
neutron star. 

%
This Alfv\'en radius ($r_{\text{A}}$) is reasonably well determined, provided we
know the neutron star's magnetic field (B$_{\text{NS}}$) and the mass-accretion
rate ($\dot{M}$). For Sco~X-1, from X-ray observations (see above) we have
estimated that $\dot{M} \simeq 0.4 - 1.2 \times 10^{-8} M_{\odot} yr^{-1}$, 
but the strength of the magnetic field is less well known. The absence of any
strong and persistent pulsation in Sco~X-1, however, and the fact that Sco~X-1 
is an LMXB system suggests that the  external magnetic field $B_{\text{NS}}$
is less than $\sim10^{12} G$. This estimate is very conservative; a more realistic 
estimate is $B_{\text{NS}} \leq 10^{10} G$ \citep{2015MNRAS.452.3994M, 
2013MNRAS.429.3411P}. In the other extreme, the minimum observed field 
values for neutron stars are $B_{\text{NS}} \simeq 10^{7} - 10^{8} G$ 
\citep{2015MNRAS.452.3994M, 2013MNRAS.429.3411P}. Considering the plausible 
uncertainties in $M$ and $M_{\odot}$ for the neutron star in Sco~X-1, this 
range corresponds to $r_{m} \simeq 10 - 2000$ km.

%
The other characteristic length-scale, the co-rotation radius ($r_{\text{co}}$) is
quantified from Kepler's 3rd Law as: 

\begin{equation}
\begin{split}
r_{\text{co}} = \bigg(\frac{GMP_{\text{spin}}^{2}}{4\pi^{2}}\bigg)^{1/3} \simeq 1.7
\times 10^{8} \times \bigg(\frac{P_{\text{spin}}}{1 s}\bigg)^{2/3} 
\times \bigg(\frac{M}{1.4M_{\odot}}\bigg)^{1/3} cm,
\end{split}
\end{equation}
where $P_{\text{spin}}$ is the stellar spin period of the neutron star. Again,
since no pulsation has yet been detected from the neutron star in Sco~X-1, its
$P_{\text{spin}}$ is completely unknown. However, as it is a highly accreting
LMXB system, its spin period is likely to be a few milliseconds, similar to
that of observed recycled pulsars. Thus in our analysis, we consider the
astrophysically plausible range of $f_{\text{spin}} (= 1/P_{\text{spin}}) \simeq 
25 -750$ Hz. Together with the assumption that $M \simeq 1.1 - 2.5 M_{\odot}$, 
the expected range of $r_{\text{co}}$ corresponds to $r_{\text{co}} \simeq 
20 - 300$ km. Therefore, the very conservative astrophysical range of
$(r_{m}/r_{\text{co}})$ for the neutron star in Sco~X-1 lies between $\simeq
0.03 - 100$. 

The amount of spin-wandering depends on the ratio of two characteristic length 
scales, $(r_{m}/r_{\text{co}})$, rather than each of them individually. Note that, 
Sco~X-1 has been observed to be a Z-type LMXB, an actively accreting neutron star 
sources with mass accretion rate persistently close to Eddington luminosity ever 
since its detection. A large $\dot M$ will increase the neutron star's spin 
frequency and thus will decrease $r_{\text{co}}$, resulting in increasing the ratio 
of $(r_{m}/r_{\text{co}})$. Therefore, large values of $(r_{m}/r_{\text{co}})$ are
strongly favored for the neutron star in Sco X-1. However, persistent mass accretion 
is strongly disfavored for $(r_{m}/r_{\text{co}}) > 1$ ~\citep{1998ApJ...501L..89B, 
Illarionov:1975ei}, contrary to the observed persistent accretion rate in Sco X-1. 
It is also worth mentioning that, the spin-wandering effect is insignificant for 
small values of $(r_{m}/r_{\text{co}})$ while potentially important to candidate 
searches for large values. The most realistic astrophysical expected scenario for 
the neutron star in Sco~X-1 corresponds to $(r_{m}/r_{\text{co}}) \sim 1$. This 
case we refer to as the ``realistic'' scenario for our following study. In this 
paper, we also explore the very ``pessimistic'' scenario corresponding to 
$(r_{m}/r_{\text{co}}) \sim 100$.}

\subsection{Estimating moment of inertia ($I_{\text{spin}}$) of the neutron star in Sco~X-1}
\label{subsec2.3}

%
The moment of inertia about the spin-axis ($I_{\text{spin}}$) also plays an
important role in our analysis, since it determines the angular acceleration
due to the torque exerted on the neutron star. We considered a physically 
plausible range of $I_{\text{spin}}$ for a neutron star including rapid spin. 
We also note that $I_{\text{spin}}$ depends on the mass{\footnote {In this 
paper, by ``mass" we refer to ``gravitational mass" for simplicity, unless 
otherwise specified.}} ($M$) and currently unknown equation-of-state (EoS) 
of neutron stars.  Moreover, $I_{\text{spin}}$ can differ depending on
whether the neutron star is rapidly spinning or in a static configuration. 

\begin{figure}[t]
\begin{center}
\hspace*{-0.5cm}
\begin{tabular}{c}
\hspace*{-0.5cm}
\includegraphics[width=0.53\textwidth,angle=0]{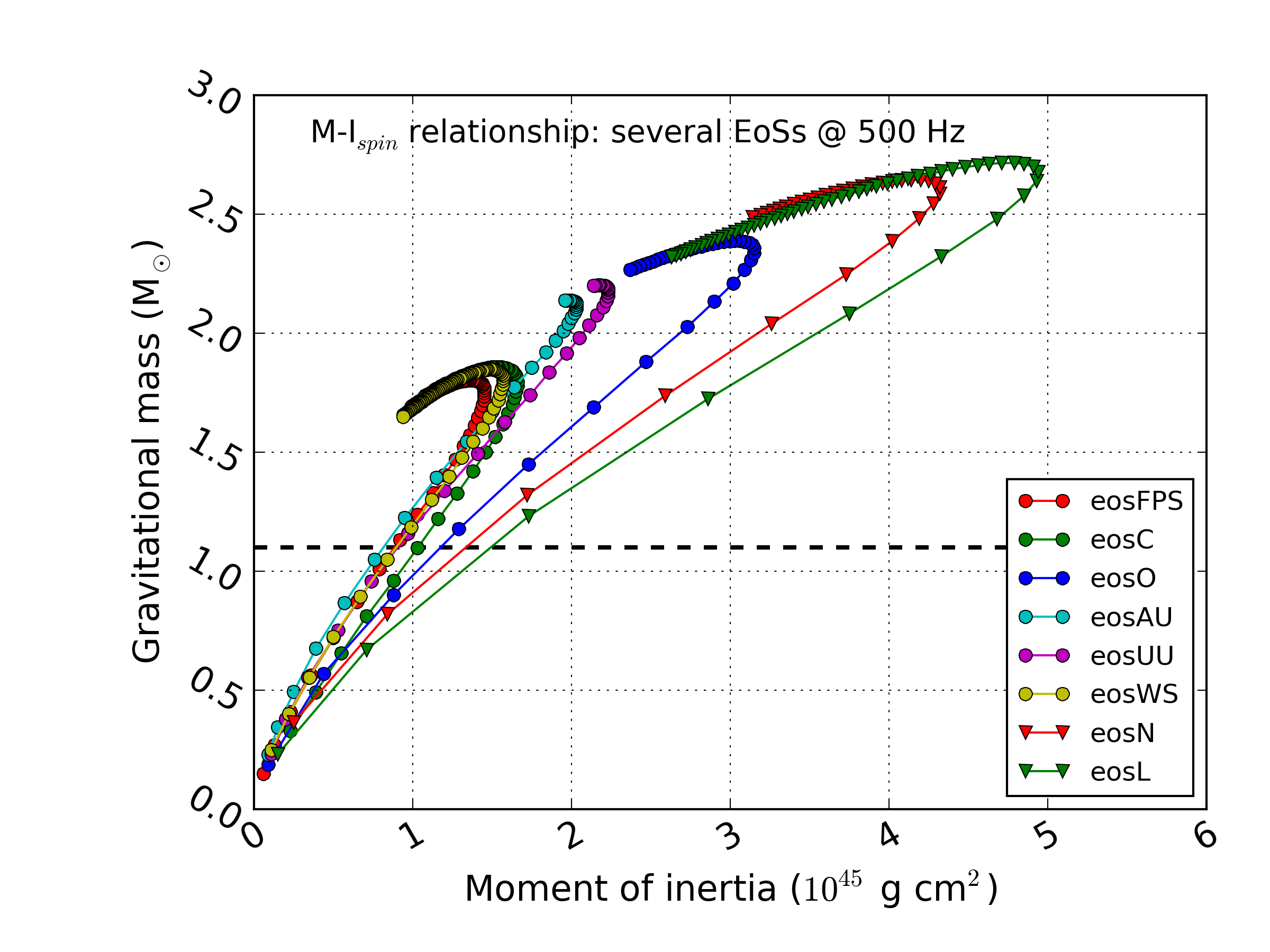}
\end{tabular}
\end{center}
\vspace*{-0.5cm}
\caption{The moment of inertia along the spin-axis ($I_{\text{spin}}$) as a function 
of different gravitational masses of neutron star for a number of realistic candidate 
equations of state is shown. Note that, $I_{\text{spin}}$ of a neutron star depends 
not only on the nature of its equations of state, but also on its gravitational mass. 
The astrophysically plausible value of gravitational mass of a neutron star lies 
somewhere between the 1.1 M$_{\odot}$ (denoted with the horizontal black dashed-line) 
and the maximum gravitational mass allowed by the particular equation of state for 
stable configuration.}
\label{M-I_Diff-EoSs}
\end{figure}

%
For this purpose, we have numerically computed the moment of inertia along the
spin-axis ($I_{\text{spin}}$) in a fully General Relativistic framework, following 
the procedure formulated in~\cite{1994ApJ...424..823C}. We have considered a number
of realistic EoSs, from soft to hard, for a neutron star over a range of stellar
spin frequencies from non-spinning up to 750 Hz. In Fig.~\ref{M-I_Diff-EoSs}, we
show the mass vs moment of inertia curves for a number of equations of state
for a stellar spin frequency of 500 Hz. The physical range of $I_{\text{spin}}$
corresponds to $\sim 0.8 - 5.0 \times 10^{45}$ g-cm$^{2}$ for different values
of mass, as well as different EoSs. 

\section{Numerical Simulation of Spin Wandering} \label{sec:level3}

\begin{figure}[h]
\begin{center}
\begin{tabular}{c}
\hspace*{-0.3cm}
\includegraphics[width=0.5\textwidth,angle=0]{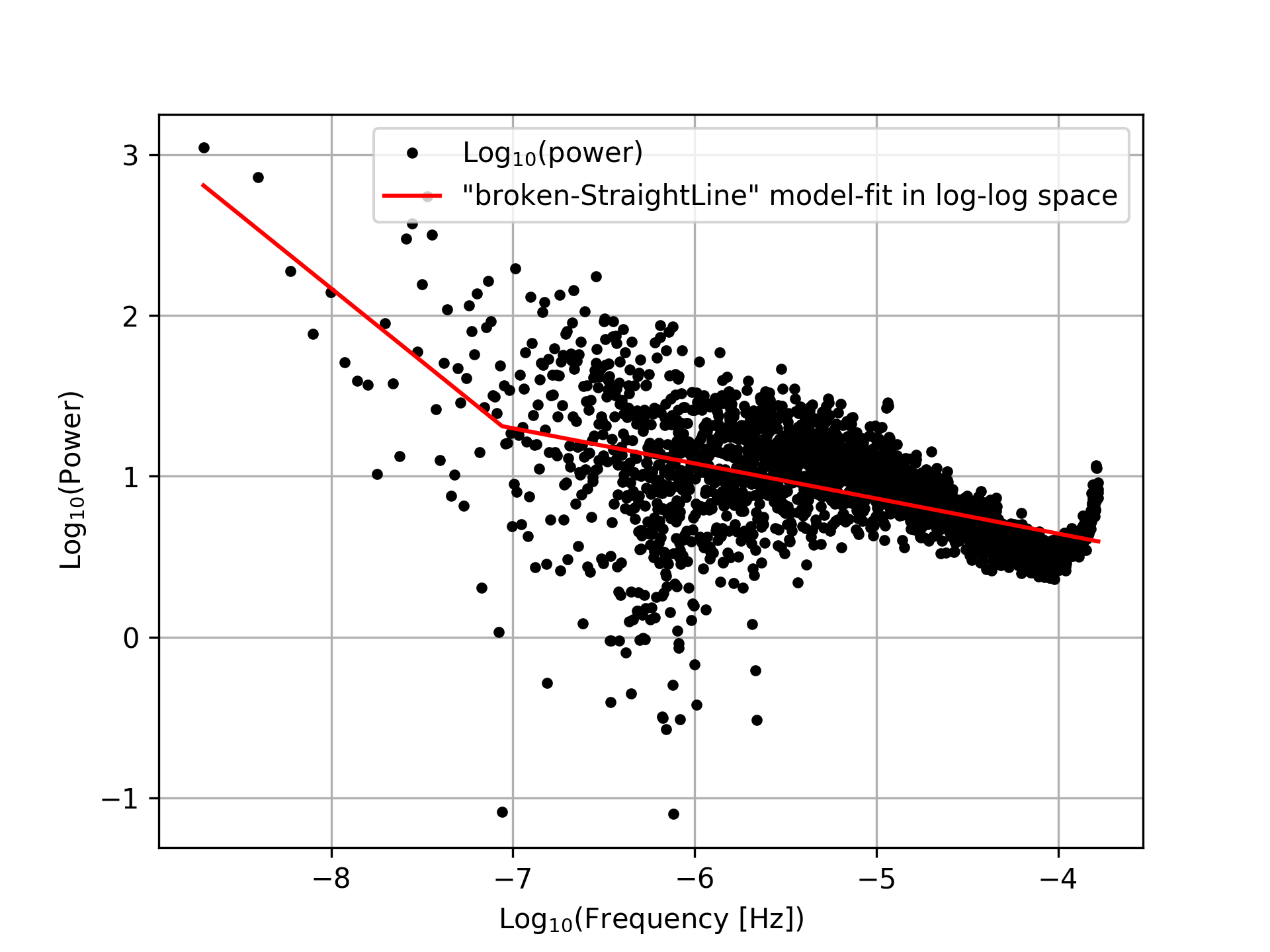}
\end{tabular}
\end{center}
\vspace*{-0.5cm}
\caption{The Lomb-Scargle periodogram computed from {\it RXTE}/ASM lightcurve of Sco~X-1 
for $\sim$ 15 year long observations. The periodogram was then fitted with a ``broken 
straight line'' model in the log-log space, representing a broken power-law model. The 
best fit model is shown with solid-red curve. We marginalize over the uncertainties in the 
estimated model parameters while performing the large-scale simulations.} 
\label{LS-periodogram}
\end{figure}

%
The purpose of this analysis is to estimate the nature and degree of this
spin-wandering behavior and to assess how continuous GW search pipelines are
potentially affected by it. For this purpose, we need to simulate a number of
spin-wandering realizations spanning the entire range of the aforementioned
astrophysically feasible parameter space. Moreover, since the spin-wandering
behavior is stochastic in nature, a statistical study over an ensemble of
realizations is more desirable than any individual realization. We briefly
describe the key steps of our simulations and analysis, below. 

%
We computed the variability of the mass accretion rate in the frequency domain
using the Lomb-Scargle periodogram\footnote{https://docs.scipy.org/doc/scipy-0.18.1/reference/\\
generated/scipy.signal.lombscargle.html}, 
since the {\it RXTE}/ASM observations of Sco X-1 were extremely unevenly spaced in 
time. The computed Lomb-Scargle periodogram was then fitted with a broken-power-law 
model as (see Fig.~\ref{LS-periodogram}) as defined by,

\begin{equation}
\begin{split}
P(f) & = A f^{-n_{1}} \qquad \qquad \forall f \leq f_{b} \\
& = A f_{b}^{n_{2} - n_{1}} f^{-n_{2}} \qquad \forall f > f_{b}
\end{split}
\end{equation}
where, $P(f)$ is the power at frequency $f$, $n_{1}$ is the power-law index 
at lower frequency $f \leq f_{b}$, $n_{2}$ is the power-law index at higher 
frequency $f > f_{b}$, and $f_{b}$ is the break frequency. The estimated 
parameter values (with their $1 \sigma$ error) of the broken-power-law model 
corresponds to $n_{1} = (0.91 \pm 0.12)$, $n_{2} = (0.22 \pm 0.074)$, 
$f_{b} = 10^{-7.06 \pm 0.10}$ Hz and $A = 10^{-5.12 \pm 0.091}
$ (count-rate)$^{2}$/Hz. It is worth mentioning that, there is an uprising 
trend in the estimated PSD at the very high-frequency end. We are not 
completely certain about its physical origin, and whether this is an 
observational/instrumental artifact. The timescale at which it tends to peak 
seems to be quite close to the {\it RXTE} orbital period ($\approx 90$ mins). 
However, this amount of power on the time-scale of $\sim$ hour has 
a negligible contribution to the time-scale of $\sim$ day or longer, where 
the spin-wandering effect is most relevant to our analysis.

%
We have simulated a number of time series realizations of the {\it RXTE}/ASM
photon count rate, applying an inverse Fourier transform of evenly-sampled
frequency series data from this broken-power-law model in the frequency domain 
randomizing over the uncertaintities in values of the estimated model parameters.
For each time series generation we randomize the phase of each Fourier frequency 
to generate a statistically random timeseries obeying the same variability
power density spectrum (PSD) as that estimated from X-ray data of {\it RXTE} 
observations. The corresponding $\dot{M}(t)$ timeseries is derived from the 
count rate time series following the calibration procedure discussed above. 
From each $\dot{M}(t)$ timeseries we compute the $\dot{f}(t)$ timeseries using 
Eqn~\ref{master-eqn}. The $\dot{f}(t)$ timeseries is then integrated with 
respect to time to obtain the instantaneous frequency timeseries $f (t)$, 
and integrated once more to obtain the instantaneous phase timeseries 
$\phi (t)$. In this integration, we choose the first integration constant 
to be the equilibrium stellar spin frequency and the second integration 
constant (reference phase) to be zero without any loss of generality.

\begin{figure}[h]
\begin{center}
\begin{tabular}{c}
\includegraphics[width=0.5\textwidth,angle=0]{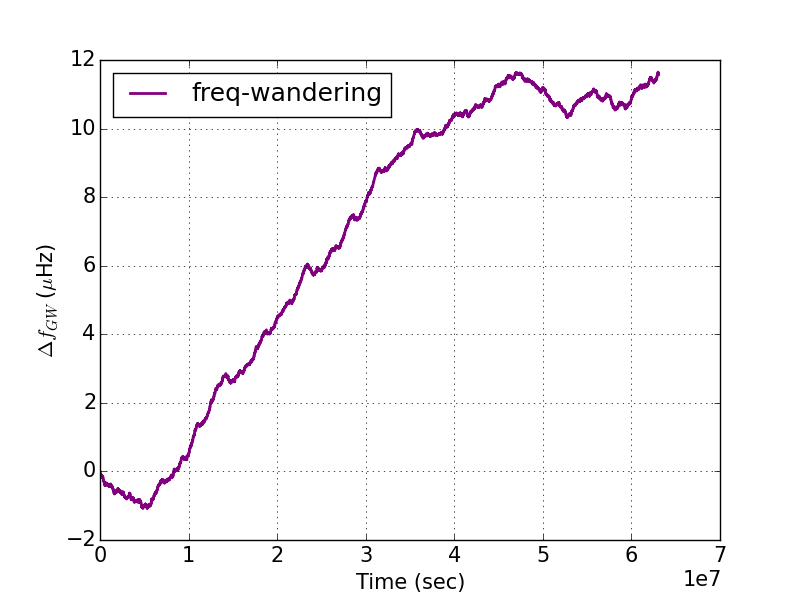} \\
\includegraphics[width=0.5\textwidth,angle=0]{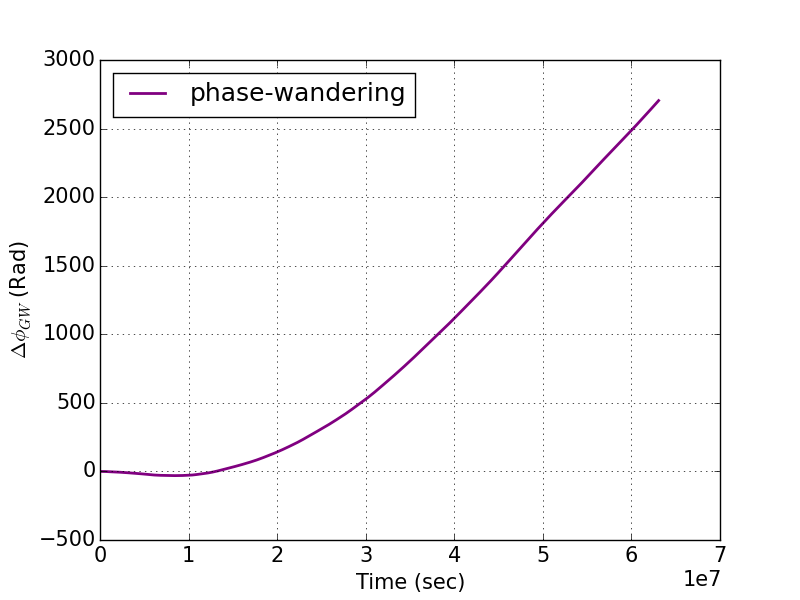}
\end{tabular}
\end{center}
\vspace*{-0.5cm}
\caption{The time evolution of a specific extreme realization of simulated spin-wandering 
timeseries for $f_{\text{spin}} = 500$ Hz, $I_{\text{spin}} = 0.8 \times 10^{45}$ g-cm$^{2}$ and 
$r_{m}/r_{\text{co}} = 100.0$. {\it {Top-panel}} demonstrates the timeseries for instantaneous 
deviation of GW-frequency (i.e., $2 \times f_{\text{spin}}$). {\it {Bottom-panel}} demonstrates 
the timeseries for cumulative deviation of GW-phase (i.e., $2 \times \phi_{\text{spin}}$).}
\label{example_SW}
\end{figure}

%
One extreme example of an $f(t)$ timeseries (termed as ``frequency wandering'') and 
the corresponding $\phi (t)$ timeseries (termed as ``phase wandering'') are shown in 
Fig.~\ref{example_SW}. This particular example corresponds to one of the realizations 
from a large number of spin-wandering simulations. In this case, the unknown system 
parameters are taken to be $f_{\text{spin}} = 500$ Hz, $I_{\text{spin}} = 0.8 \times 
10^{45}$ g-cm$^{2}$ and $r_{m}/r_{\text{co}} = 100.0$, corresponding to the above 
conservative scenario. From this example, it is clear that the phase can have an 
off-set of $\Delta \phi_{\text{GW}} \approx 1$ radian in a time-scale of a single day 
in such an extreme case. We describe our extensive simulations of the spin-wandering 
effect over several astrophysically possible ranges of unknown parameters for the 
neutron star in Sco~X-1. 

\subsection{Statistical behavior of spin-wandering of the neutron star in Sco~X-1}
\label{subsec3.1}

%
Since accretion-induced spin-wandering is stochastic, its effects are best
studied statistically. Moreover, the values of several important quantities,
such as spin frequency ($f_{\text{spin}}$), the ratio of magnetic radius to
co-rotation radius ($r_{m}/r_{\text{co}}$), and moment of inertia along the
spin-axis ($I_{\text{spin}}$), are unknown for Sco~X-1 over broad ranges.

\begin{figure*}
\begin{center}
\hspace*{-0.75cm}
\begin{tabular}{c}
\includegraphics[width=\textwidth,angle=0]{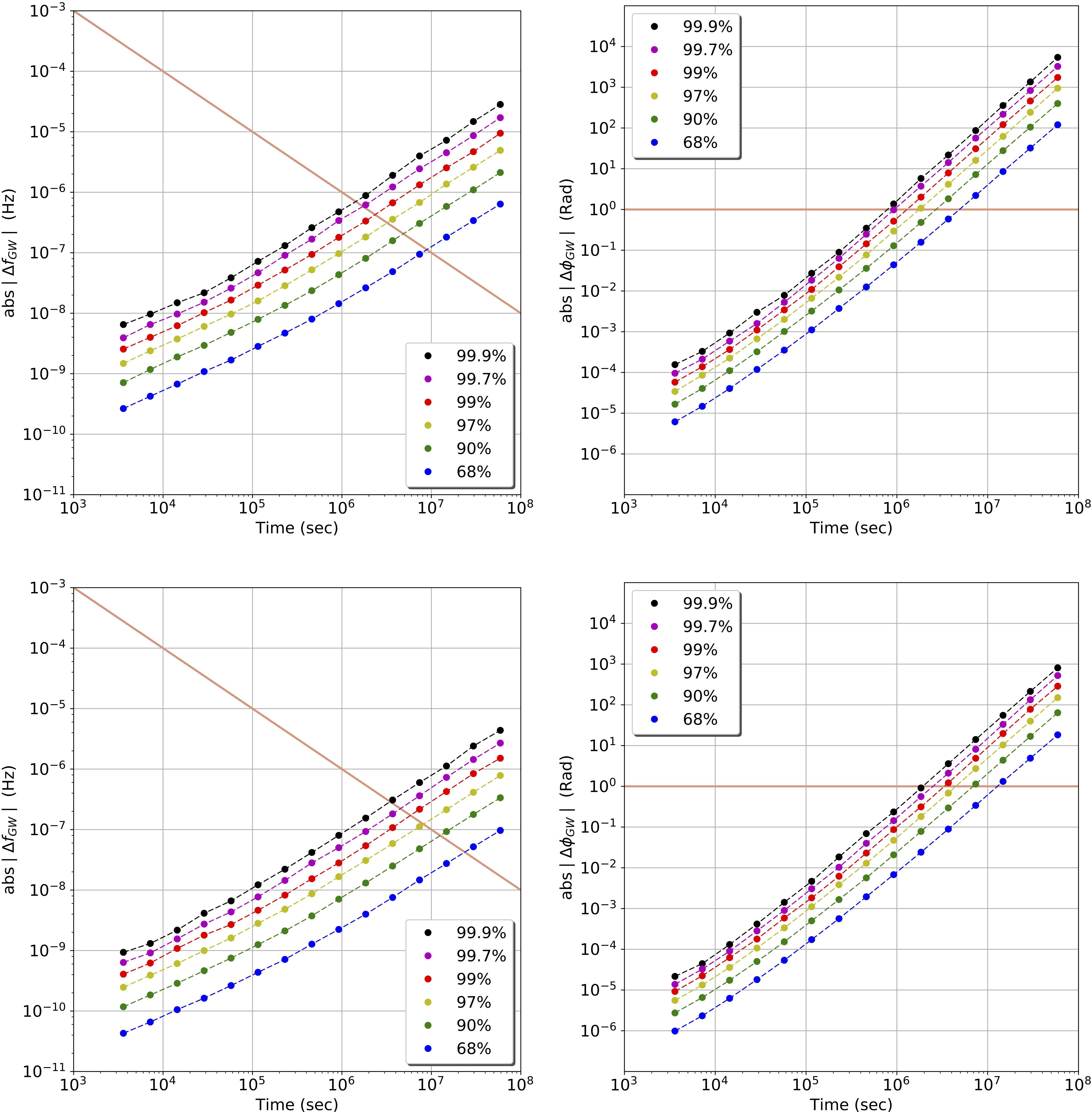} \\
\end{tabular}
\end{center}
\caption{This figure quantifies the statistical estimate of spin-wandering effect 
for the neutron star in Sco~X-1. In each sub-plot different values of statistical 
confidence levels (see legends) are plotted that denote the cumulative probability of 
maximum frequency-wandering ({\it left-column}) and phase-wandering ({\it right-column}) 
effects. In this figure we take the unknown physical parameters of $f_{\text{spin}} = 
25$ Hz and $r_{m}/r_{\text{co}} = 1$ (the realistic scenario); 
corresponding to 2,2-mode GW-frequency $f_{\text{GW}} = 50$ Hz. The {\it top-row} 
corresponds to the minimum theoretical value of moment of inertia, $I_{\text{spin}} = 
0.8 \times 10^{45}$ g-cm$^{2}$, while the {\it bottom-row} corresponds to the maximum 
theoretical value of moment of inertia $I_{\text{spin}} = 5 \times 10^{45}$ g-cm$^{2}$. 
For each case of a specific choices of $f_{\text{spin}}$, $(r_{m}/r_{\text{co}})$ and 
$I_{\text{spin}}$, we simulated 10,000 spin-wandering realizations and estimated the 
amount of deviation in instantaneous $f_{\text{GW}}$ and $\phi_{\text{GW}}$ from their 
initial values at a number of statistical significance levels, 68\%, 90\%, 97\%, 99\%, 
99.7\% and 99.9\% as a function of time. The absolute value of deviation in 
$f_{\text{GW}}$ and $\phi_{\text{GW}}$ for a given significance-level corresponds 
to the fraction of the spin-wandering realizations remaining within the respective 
range. Each of the orange straight-lines for frequency-wandering cases (panels in 
the left-column) denotes the frequency resolution as a function of time of integration 
of the signal. The orange straight-lines for phase-wandering cases (panels in the 
right-column) denote phase mismatches of 1 rad due to the spin-wandering effect.} 
\label{realistic-50Hz_SWstatPlot}
\end{figure*}

\begin{figure*}
\begin{center}
\hspace*{-0.75cm}
\begin{tabular}{c}
\includegraphics[width=\textwidth,angle=0]{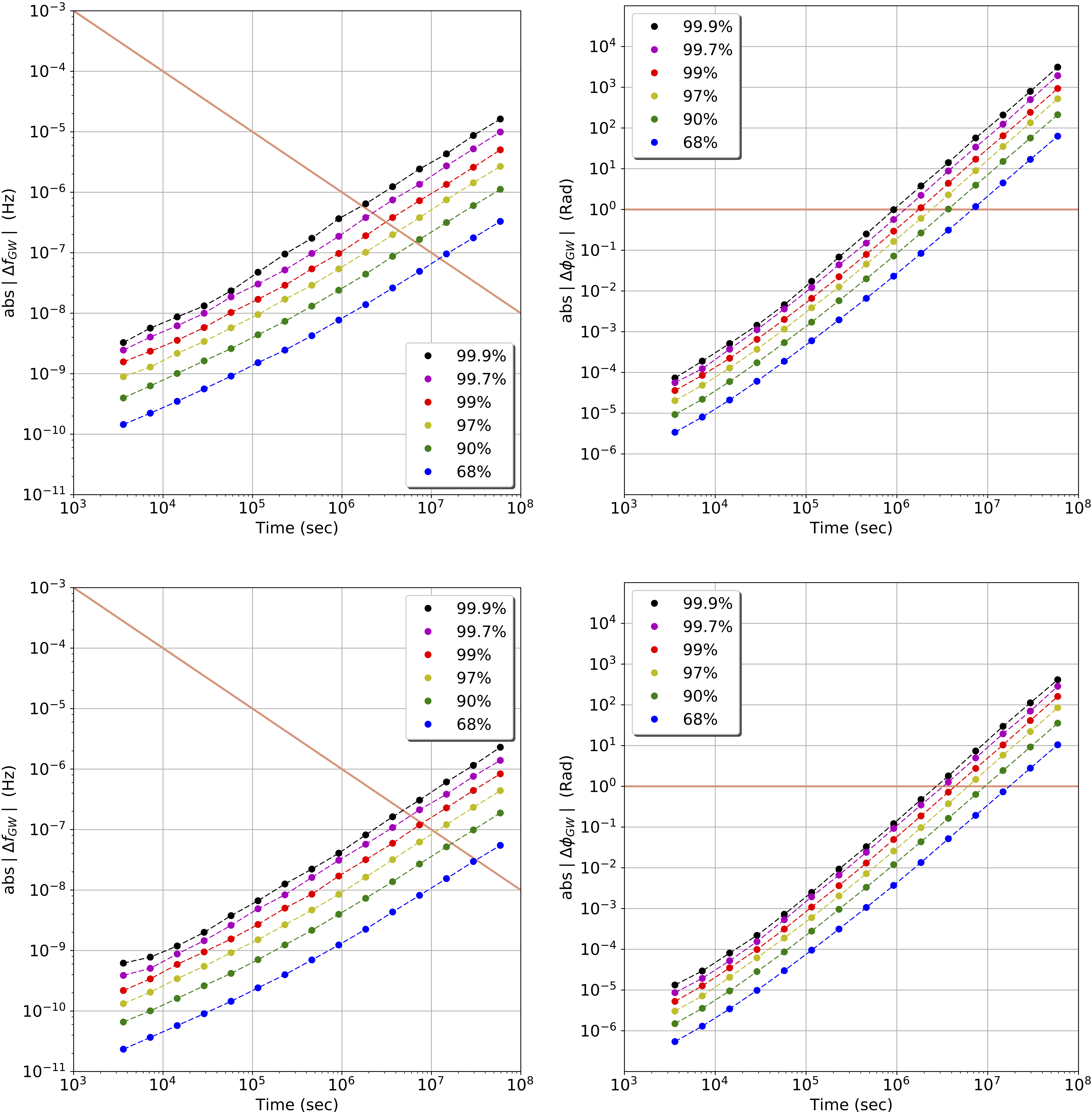} \\
\end{tabular}
\end{center}
\caption{This figure is similar to Figure~\ref{realistic-50Hz_SWstatPlot}, but with 
different values of the unknown physical parameters. In this figure we take parameters 
of $f_{\text{spin}} = 150$ Hz and $r_{m}/r_{\text{co}} = 1$ (the realistic scenario); 
corresponding to 2,2-mode GW-frequency $f_{\text{GW}} = 300$ Hz. 
The {\it top-row} corresponds to the minimum theoretical value of moment of inertia, 
$I_{\text{spin}} = 0.8 \times 10^{45}$ g-cm$^{2}$, while the {\it bottom-row} 
corresponds to the maximum theoretical value of moment of inertia $I_{\text{spin}} 
= 5 \times 10^{45}$ g-cm$^{2}$.} 
\label{realistic-300Hz_SWstatPlot}
\end{figure*}

\begin{figure*}
\begin{center}
\hspace*{-0.75cm}
\begin{tabular}{c}
\includegraphics[width=\textwidth,angle=0]{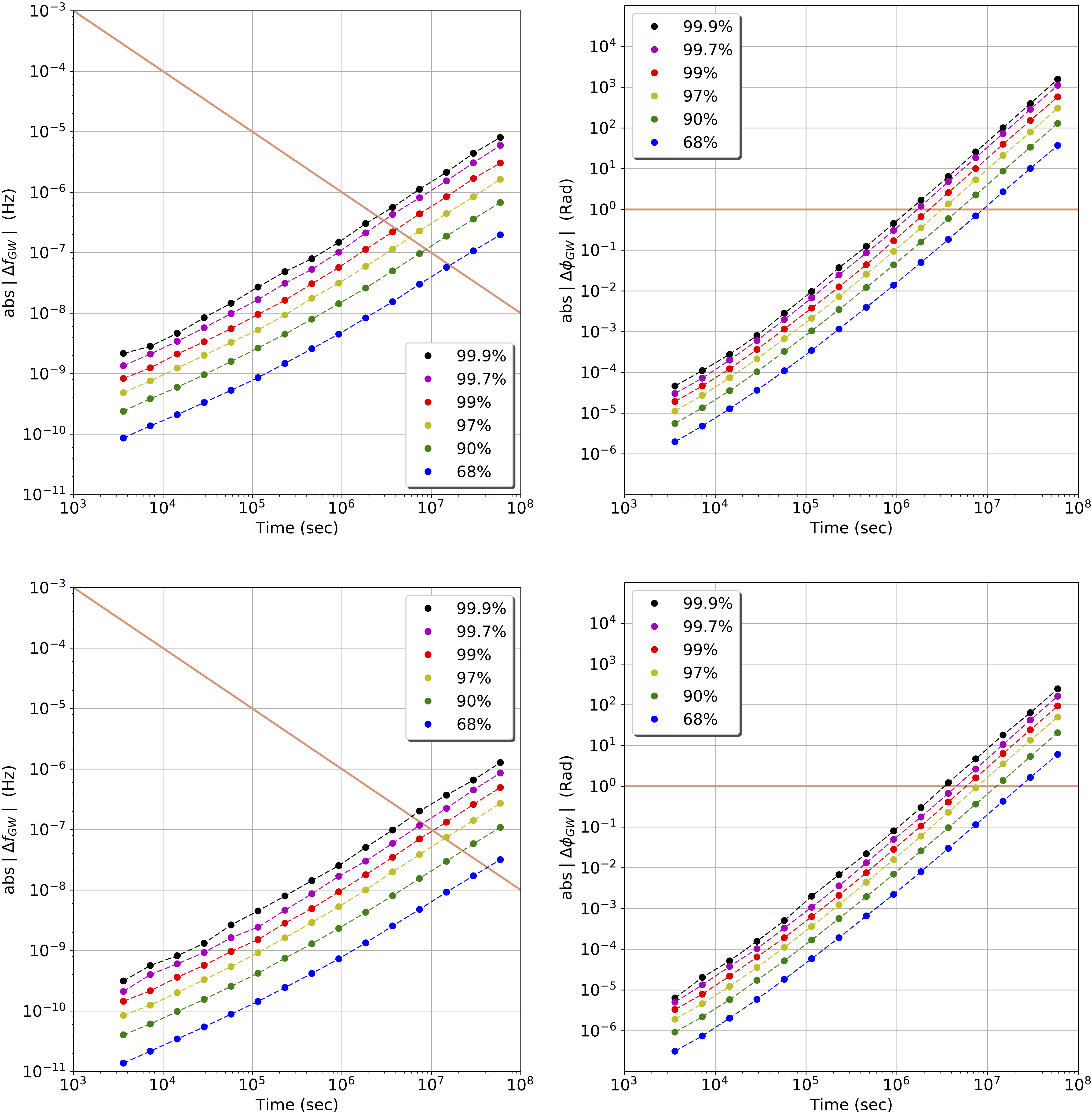} \\
\end{tabular}
\end{center}
\caption{This figure is similar to Figure~\ref{realistic-50Hz_SWstatPlot}, but with 
different values of the unknown physical parameters. In this figure we take parameters 
of $f_{\text{spin}} = 750$ Hz and $r_{m}/r_{\text{co}} = 1$ (the realistic scenario); 
corresponding to 2,2-mode GW-frequency $f_{\text{GW}} = 1500$ Hz. 
The {\it top-row} corresponds to the minimum theoretical value of moment of inertia, 
$I_{\text{spin}} = 0.8 \times 10^{45}$ g-cm$^{2}$, while the {\it bottom-row} corresponds 
to the maximum theoretical value of moment of inertia $I_{\text{spin}} = 5 \times 
10^{45}$ g-cm$^{2}$.} 
\label{realistic-1500Hz_SWstatPlot}
\end{figure*}

\begin{figure*}
\begin{center}
\hspace*{-0.75cm}
\begin{tabular}{c}
\includegraphics[width=\textwidth,angle=0]{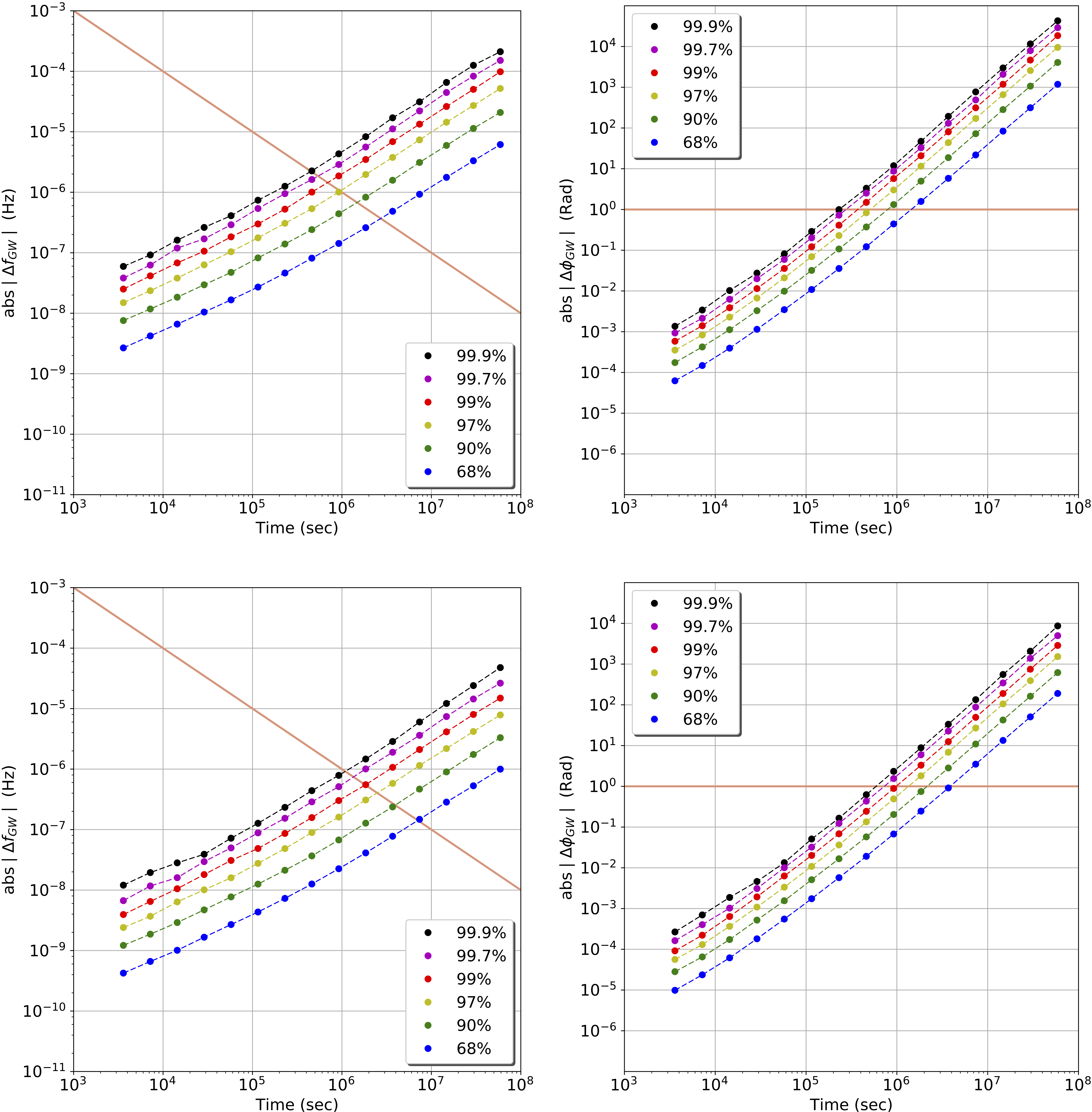} \\
\end{tabular}
\end{center}
\caption{This figure is similar to Figure~\ref{realistic-50Hz_SWstatPlot}, but with different 
values of the unknown physical parameters. In this figure we take parameters of $f_{\text{spin}} 
= 25$ Hz and $r_{m}/r_{\text{co}} = 100$ (the pessimistic scenario); corresponding to 
2,2-mode GW-frequency $f_{\text{GW}} = 50$ Hz. The {\it top-row} corresponds to 
the minimum theoretical value of moment of inertia, $I_{\text{spin}} = 0.8 \times 10^{45}$ 
g-cm$^{2}$, while the {\it bottom-row} corresponds to the maximum theoretical value of 
moment of inertia $I_{\text{spin}} = 5 \times 10^{45}$ g-cm$^{2}$.} 
\label{conservative-50Hz_SWstatPlot}
\end{figure*}

\begin{figure*}
\begin{center}
\hspace*{-0.75cm}
\begin{tabular}{c}
\includegraphics[width=\textwidth,angle=0]{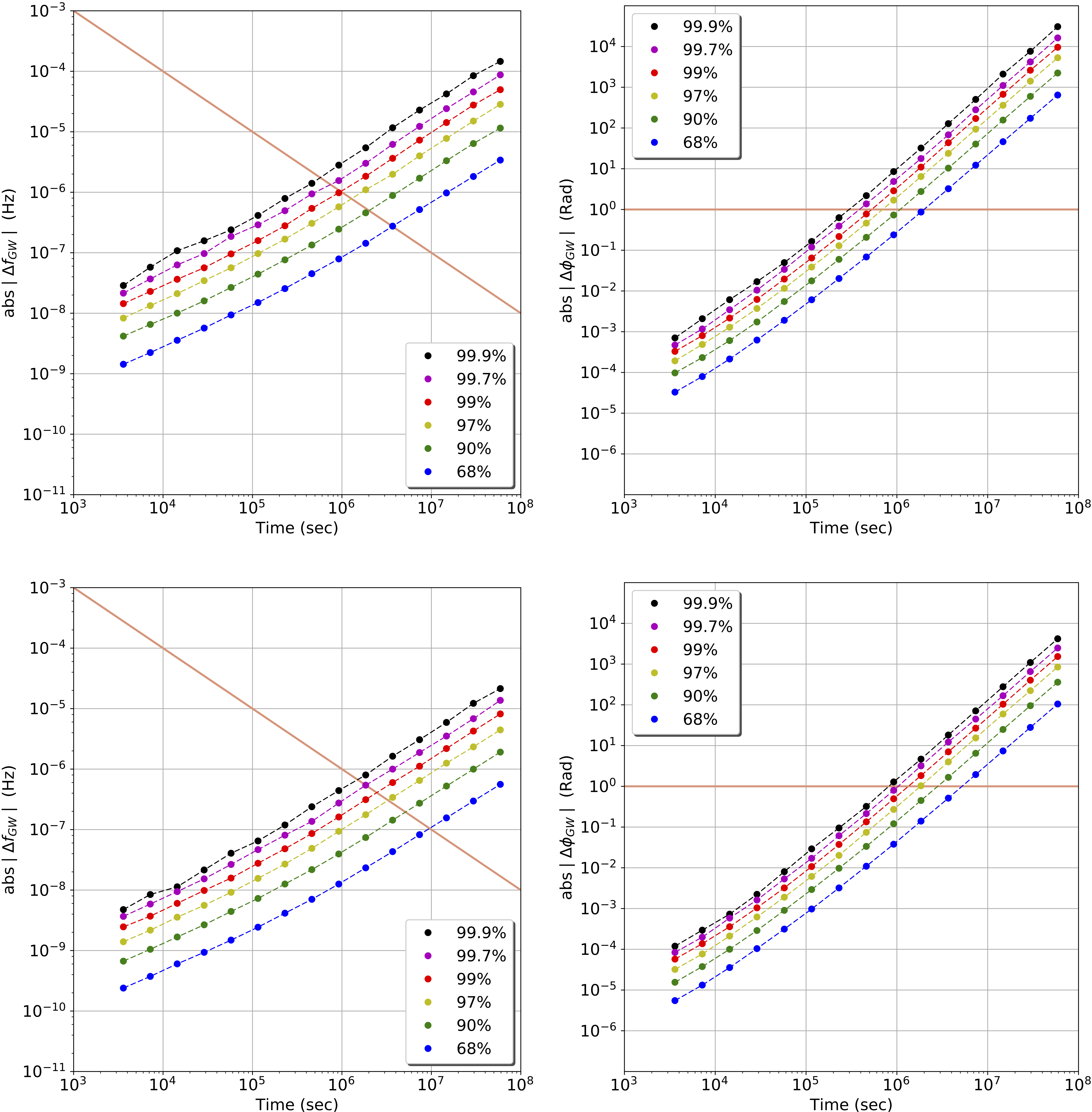} \\
\end{tabular}
\end{center}
\caption{This figure is similar to Figure~\ref{realistic-50Hz_SWstatPlot}, but with different 
values of the unknown physical parameters. In this figure we take parameters of $f_{\text{spin}} 
= 150$ Hz and $r_{m}/r_{\text{co}} = 100$ (the pessimistic scenario); 
corresponding to 2,2-mode GW-frequency $f_{\text{GW}} = 300$ Hz. The {\it top-row} corresponds to 
the minimum theoretical value of moment of inertia, $I_{\text{spin}} = 0.8 \times 10^{45}$ 
g-cm$^{2}$, while the {\it bottom-row} corresponds to the maximum theoretical value of 
moment of inertia $I_{\text{spin}} = 5 \times 10^{45}$ g-cm$^{2}$.} 
\label{conservative-300Hz_SWstatPlot}
\end{figure*}

\begin{figure*}
\begin{center}
\hspace*{-0.75cm}
\begin{tabular}{c}
\includegraphics[width=\textwidth,angle=0]{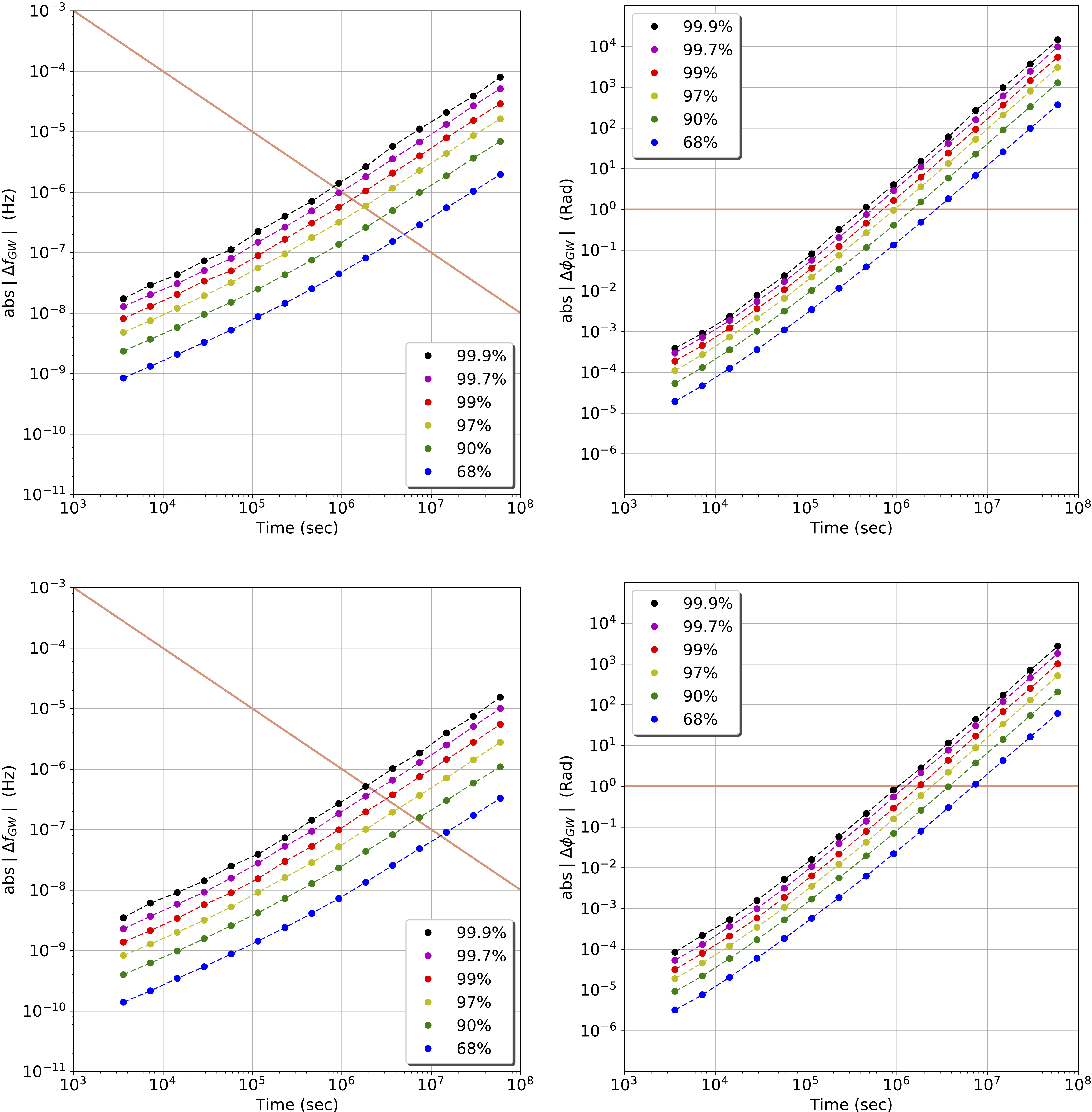} \\
\end{tabular}
\end{center}
\caption{This figure is similar to Figure~\ref{realistic-50Hz_SWstatPlot}, but with different 
values of the unknown physical parameters. In this figure we take parameters of $f_{\text{spin}} 
= 750$ Hz and $r_{m}/r_{co} = 100$ (the pessimistic scenario); 
corresponding to 2,2-mode GW-frequency $f_{\text{GW}} = 1500$ Hz. The {\it top-row} corresponds 
to the minimum theoretical value of moment of inertia, $I_{\text{spin}} = 0.8 \times 10^{45}$ 
g-cm$^{2}$, while the {\it bottom-row} corresponds to the maximum theoretical value of 
moment of inertia $I_{\text{spin}} = 5 \times 10^{45}$ g-cm$^{2}$.} 
\label{conservative-1500Hz_SWstatPlot}
\end{figure*}

%
For this purpose, we have simulated several cases of spin-wandering effects, 
each case with a different set of astrophysical parameters, $f_{\text{spin}}$, 
$(r_{m}/r_{\text{co}})$ and $I_{\text{spin}}$. We have performed a large number of 
simulations for each set of assumed parameter values spanning their astrophysical 
expected range for the neutron star in Sco~X-1. For each case of a specific 
choices of $f_{\text{spin}}$, $(r_{m}/r_{\text{co}})$ and $I_{\text{spin}}$, we have
simulated 10,000 spin-wandering realizations and estimated the amount of deviation 
in instantaneous $f_{\text{GW}}$ and $\phi_{\text{GW}}$ from their initial values 
at a number of statistical significance levels, 68\%, 90\%, 97\%, 99\%, 99.7\% and 
99.9\% as a function of time in the figures (Figure~\ref{realistic-50Hz_SWstatPlot} 
to Figure~\ref{conservative-1500Hz_SWstatPlot}). The absolute value of deviation in 
$f_{\text{GW}}$ and $\phi_{\text{GW}}$ for a given significance-level corresponds 
to the fraction of the spin-wandering realizations remaining within the respective 
range. 

We broadly classify all the simulations into two cases, the first one corresponding 
to $r_{m}/r_{\text{co}} = 1$, being the more realistic scenario. Next, we consider 
the extreme and pessimistic scenario corresponding to $r_{m}/r_{\text{co}} = 100$, 
leading to the maximum amount of spin-wandering effect for this source. We show 
the results of this large scale simulations in the Figures~\ref{realistic-50Hz_SWstatPlot}, 
\ref{realistic-300Hz_SWstatPlot}, \ref{realistic-1500Hz_SWstatPlot} for 
the realistic scenario, and Figures~\ref{conservative-50Hz_SWstatPlot}, 
\ref{conservative-300Hz_SWstatPlot}, \ref{conservative-1500Hz_SWstatPlot}, 
for the pessimistic scenario. 

In these figures, we demonstrate both ``frequency-wandering'' and ``phase-wandering'' 
behavior quantitatively in statistical measure. We performed simulations for three 
values of continuous gravitational wave frequencies ($f_{\text{GW}}$), (i) at the 
low-frequency end (50 Hz), (ii) at a mid-frequency where the Advanced-LIGO and 
Advanced-Virgo noise sensitivity is good (300 Hz), and (iii) at the high-frequency 
end (1500 Hz), for each of the two scenarios. Moreover, for each of the 
aforementioned cases, we perform our simulations for the two extreme values of 
theoretically feasible moment of inertia along the spin-axis, $I_{\text{spin}} = 0.8 
\times 10^{45}$ g-cm$^{2}$ and $I_{\text{spin}} = 5 \times 10^{45}$ g-cm$^{2}$, for the 
neutron star in Sco~X-1. 

\section{Conclusion and Discussion} \label{sec:level4}

%
In this paper, we studied the spin-wandering effect of the neutron star in 
Sco~X-1 using the X-ray observations (publicly available archival data from 
{\tt HEASARC} facility of GSFC/NASA and MIT/XTE team) of the source with ASM and 
PCA instruments onboard the {\it RXTE} satellite. We developed and implemented 
the first methodology to infer the frequency wandering and phase wandering, 
collectively termed as ``the spin-wandering effect'', quantitatively with a 
range of astrophysically feasible parameter ranges. We have also studied the 
statistical behaviors of both frequency wandering and phase wandering. 

The purpose of this study is not to perform a precise estimation of the 
spin-wandering effect nor to develop an improved theoretical model of accretion 
induced torque on the neutron star of an accreting LMXB system. Instead, we 
applied the existing standard physical model of accretion-induced torque by 
\citet{1997ApJS..113..367B} on the accreting neutron star of Sco~X-1 to infer 
the astrophysically expected range of the spin wandering.

%
We found the estimated effects of spin wandering can vary greatly, depending on 
the assumed astrophysical parameters of Sco~X-1. In many cases, however,
the degree of spin-wandering is significant and can plausibly degrade the
sensitivity of some continuous gravitational wave search pipelines,
particularly those that gain sensitivity from long coherence times and make
rigid assumptions about the source frequency evolution. As a consequence, it is
important for GW search pipelines to evaluate their realistic sensitivity given
the full range of astrophysically plausible spin-wandering effects presented in
this paper. 

%
The results of the first mock-data-challenge (MDC) specifically designed to
test GW search pipelines in their ability to detect signals from Sco~X-1 were
published in~\cite{2015PhRvD..92b3006M}. This analysis did not explicitly
include spin-wandering effects in the mock data analysed by the challenge
participants.  We plan to use the spin-wandering simulation code developed 
for this paper in the generation of data for the second MDC. This will
add an additional level of realism to the data and enable existing search
pipelines to demonstrate their robustness against differing degrees of spin
wandering.

%
Our analysis and modelling have some limitations, and a number of improvements
could be made in the event of futher relevant EM observations of Sco~X-1. Here,
we briefly mention a few key issues. First, we highlight our ignorance of the
nominal spin frequency of Sco~X-1. A future detection of coherent pulsations or
burst oscillations would be immensely useful in narrowing down the uncertainty
in several ways, including not only the approximate frequency band, but also in
the estimation of the co-rotation radius ($r_{\text{co}}$), which affects the 
degree of spin wandering.

%
Any determination of the magnetic field of the neutron star in Sco~X-1 would
also be useful for estimating the magnetic radius ($r_{m}$) and thus allow us
to better estimate the degree of expected spin wandering. Moreover, improved
knowledge of the mass, and the correct EoS of the neutron star would also help
improve the estimation of $I_{\text{spin}}$, which also affects the degree of
spin wandering. Finally, note that we have ignored any possible ``propeller effect" 
that may arise in disk-magnetospheric interaction in our analysis.

%
In future, long duration X-ray observations, preferably at equally-sampled
time intervals would be very useful. In our analysis, we used the publicly
available {\it RXTE}/ASM data of Sco~X-1 from the MIT/XTE team and the {\tt
HEASARC} facility of GSFC/NASA. The available observational data are extremely
sparsely sampled, however. This prevented use of conventional Fourier transform
methods to estimate the variability characteristics in the frequency domain.
Instead, we used Lomb-Scargle periodograms which are subject to larger spectral
leakage, leading to potential systematic bias in our analysis. Present and
upcoming X-ray instruments, e.g., the Scanning Sky Monitor (SSM) onboard {\it
AstroSat}, could be very useful, particularly the event mode data, containing 
information of individual photon arrival time, to mitigate this problem. 

\maketitle

\paragraph{Acknowledgments:---} 

This research has made use of data and software provided by the High Energy 
Astrophysics Science Archive Research Center ({\tt HEASARC}), which is a service 
of the Astrophysics Science Division at NASA/GSFC and the High Energy Astrophysics 
Division of the Smithsonian Astrophysical Observatory. A.~M.~thanks Alan Levine, 
Tod Strohmayer and XTE team at MIT for sharing information on the {\it RXTE}/ASM 
data. The authors would also like to acknowledge valuable input from several members 
within LIGO Scientific Collaboration (LSC) and Virgo collaboration, and would particularly 
like to thank G.~Ashton, G.~Meadors, L.~Sun, and J.~T.~Whelan for their comments 
on an early draft. A.~M.~acknowledges support from the “SERB Start-Up Research for
Young Scientists Scheme” project Grant No. SB/FTP/ PS-067/2014, DST, India.
K.~R.~acknowledges support from the National Science Foundation (Grant
PHY-1505932). Most of the computational studies reported here were run on the
ATLAS computing cluster at {\it Max-Planck-Institut f{\"u}r Gravitationsphysik}
(Albert-Einstein-Institut), Hannover and computational cluster Mowgli
International Centre for Theoretical Sciences, TIFR, Bangalore. The authors
gratefully acknowledge the support of the the Max Planck Society and the State
of Niedersachsen, Germany, for provision of computational resources. This paper
has LIGO document number {\tt LIGO-P1700190}.

\section*{References}

%
%
\bibliography{masterbib}

\begin{thebibliography}{24}%
\makeatletter
\providecommand \@ifxundefined [1]{%
 \@ifx{#1\undefined}
}%
\providecommand \@ifnum [1]{%
 \ifnum #1\expandafter \@firstoftwo
 \else \expandafter \@secondoftwo
 \fi
}%
\providecommand \@ifx [1]{%
 \ifx #1\expandafter \@firstoftwo
 \else \expandafter \@secondoftwo
 \fi
}%
\providecommand \natexlab [1]{#1}%
\providecommand \enquote  [1]{``#1''}%
\providecommand \bibnamefont  [1]{#1}%
\providecommand \bibfnamefont [1]{#1}%
\providecommand \citenamefont [1]{#1}%
\providecommand \href@noop [0]{\@secondoftwo}%
\providecommand \href [0]{\begingroup \@sanitize@url \@href}%
\providecommand \@href[1]{\@@startlink{#1}\@@href}%
\providecommand \@@href[1]{\endgroup#1\@@endlink}%
\providecommand \@sanitize@url [0]{\catcode `\\12\catcode `\$12\catcode
  `\&12\catcode `\#12\catcode `\^12\catcode `\_12\catcode `\%12\relax}%
\providecommand \@@startlink[1]{}%
\providecommand \@@endlink[0]{}%
\providecommand \url  [0]{\begingroup\@sanitize@url \@url }%
\providecommand \@url [1]{\endgroup\@href {#1}{\urlprefix }}%
\providecommand \urlprefix  [0]{URL }%
\providecommand \Eprint [0]{\href }%
\providecommand \doibase [0]{http://dx.doi.org/}%
\providecommand \selectlanguage [0]{\@gobble}%
\providecommand \bibinfo  [0]{\@secondoftwo}%
\providecommand \bibfield  [0]{\@secondoftwo}%
\providecommand \translation [1]{[#1]}%
\providecommand \BibitemOpen [0]{}%
\providecommand \bibitemStop [0]{}%
\providecommand \bibitemNoStop [0]{.\EOS\space}%
\providecommand \EOS [0]{\spacefactor3000\relax}%
\providecommand \BibitemShut  [1]{\csname bibitem#1\endcsname}%
\let\auto@bib@innerbib\@empty
\bibitem [{\citenamefont {{Bildsten}}(1998)}]{1998ApJ...501L..89B}%
  \BibitemOpen
  \bibfield  {author} {\bibinfo {author} {\bibfnamefont {L.}~\bibnamefont
  {{Bildsten}}},\ }\href {\doibase 10.1086/311440} {\bibfield  {journal}
  {\bibinfo  {journal} {\apjl}\ }\textbf {\bibinfo {volume} {501}},\ \bibinfo
  {pages} {L89} (\bibinfo {year} {1998})},\ \Eprint
  {http://arxiv.org/abs/astro-ph/9804325} {astro-ph/9804325} \BibitemShut
  {NoStop}%
\bibitem [{\citenamefont {{Bildsten}}\ \emph {et~al.}(1997)\citenamefont
  {{Bildsten}}, \citenamefont {{Chakrabarty}}, \citenamefont {{Chiu}},
  \citenamefont {{Finger}}, \citenamefont {{Koh}}, \citenamefont {{Nelson}},
  \citenamefont {{Prince}}, \citenamefont {{Rubin}}, \citenamefont {{Scott}},
  \citenamefont {{Stollberg}}, \citenamefont {{Vaughan}}, \citenamefont
  {{Wilson}},\ and\ \citenamefont {{Wilson}}}]{1997ApJS..113..367B}%
  \BibitemOpen
  \bibfield  {author} {\bibinfo {author} {\bibfnamefont {L.}~\bibnamefont
  {{Bildsten}}}, \bibinfo {author} {\bibfnamefont {D.}~\bibnamefont
  {{Chakrabarty}}}, \bibinfo {author} {\bibfnamefont {J.}~\bibnamefont
  {{Chiu}}}, \bibinfo {author} {\bibfnamefont {M.~H.}\ \bibnamefont
  {{Finger}}}, \bibinfo {author} {\bibfnamefont {D.~T.}\ \bibnamefont {{Koh}}},
  \bibinfo {author} {\bibfnamefont {R.~W.}\ \bibnamefont {{Nelson}}}, \bibinfo
  {author} {\bibfnamefont {T.~A.}\ \bibnamefont {{Prince}}}, \bibinfo {author}
  {\bibfnamefont {B.~C.}\ \bibnamefont {{Rubin}}}, \bibinfo {author}
  {\bibfnamefont {D.~M.}\ \bibnamefont {{Scott}}}, \bibinfo {author}
  {\bibfnamefont {M.}~\bibnamefont {{Stollberg}}}, \bibinfo {author}
  {\bibfnamefont {B.~A.}\ \bibnamefont {{Vaughan}}}, \bibinfo {author}
  {\bibfnamefont {C.~A.}\ \bibnamefont {{Wilson}}}, \ and\ \bibinfo {author}
  {\bibfnamefont {R.~B.}\ \bibnamefont {{Wilson}}},\ }\href {\doibase
  10.1086/313060} {\bibfield  {journal} {\bibinfo  {journal} {\apjs}\ }\textbf
  {\bibinfo {volume} {113}},\ \bibinfo {pages} {367} (\bibinfo {year}
  {1997})},\ \Eprint {http://arxiv.org/abs/astro-ph/9707125} {astro-ph/9707125}
  \BibitemShut {NoStop}%
\bibitem [{\citenamefont {Chakrabarty}(2008)}]{2008AIPC.1068...67C}%
  \BibitemOpen
  \bibfield  {author} {\bibinfo {author} {\bibfnamefont {D.}~\bibnamefont
  {Chakrabarty}},\ }\href@noop {} {\bibfield  {journal} {\bibinfo  {journal}
  {arXiv.org}\ ,\ \bibinfo {pages} {67}} (\bibinfo {year} {2008})},\ \Eprint
  {http://arxiv.org/abs/0809.4031v1} {0809.4031v1} \BibitemShut {NoStop}%
\bibitem [{\citenamefont {{Patruno}}\ \emph {et~al.}(2017)\citenamefont
  {{Patruno}}, \citenamefont {{Haskell}},\ and\ \citenamefont
  {{Andersson}}}]{2017arXiv170507669P}%
  \BibitemOpen
  \bibfield  {author} {\bibinfo {author} {\bibfnamefont {A.}~\bibnamefont
  {{Patruno}}}, \bibinfo {author} {\bibfnamefont {B.}~\bibnamefont
  {{Haskell}}}, \ and\ \bibinfo {author} {\bibfnamefont {N.}~\bibnamefont
  {{Andersson}}},\ }\href@noop {} {\bibfield  {journal} {\bibinfo  {journal}
  {ArXiv e-prints}\ } (\bibinfo {year} {2017})},\ \Eprint
  {http://arxiv.org/abs/1705.07669} {arXiv:1705.07669 [astro-ph.HE]}
  \BibitemShut {NoStop}%
\bibitem [{\citenamefont {Abbott}\ \emph
  {et~al.}(2016{\natexlab{a}})\citenamefont {Abbott} \emph
  {et~al.}}]{Abbott:2016blz}%
  \BibitemOpen
  \bibfield  {author} {\bibinfo {author} {\bibfnamefont {B.~P.}\ \bibnamefont
  {Abbott}} \emph {et~al.} (\bibinfo {collaboration} {Virgo, LIGO
  Scientific}),\ }\href {\doibase 10.1103/PhysRevLett.116.061102} {\bibfield
  {journal} {\bibinfo  {journal} {Phys. Rev. Lett.}\ }\textbf {\bibinfo
  {volume} {116}},\ \bibinfo {pages} {061102} (\bibinfo {year}
  {2016}{\natexlab{a}})},\ \Eprint {http://arxiv.org/abs/1602.03837}
  {arXiv:1602.03837 [gr-qc]} \BibitemShut {NoStop}%
\bibitem [{\citenamefont {Abbott}\ \emph
  {et~al.}(2016{\natexlab{b}})\citenamefont {Abbott} \emph
  {et~al.}}]{Abbott:2016nmj}%
  \BibitemOpen
  \bibfield  {author} {\bibinfo {author} {\bibfnamefont {B.~P.}\ \bibnamefont
  {Abbott}} \emph {et~al.} (\bibinfo {collaboration} {Virgo, LIGO
  Scientific}),\ }\href {\doibase 10.1103/PhysRevLett.116.241103} {\bibfield
  {journal} {\bibinfo  {journal} {Phys. Rev. Lett.}\ }\textbf {\bibinfo
  {volume} {116}},\ \bibinfo {pages} {241103} (\bibinfo {year}
  {2016}{\natexlab{b}})},\ \Eprint {http://arxiv.org/abs/1606.04855}
  {arXiv:1606.04855 [gr-qc]} \BibitemShut {NoStop}%
\bibitem [{\citenamefont {{Abbott}}\ \emph {et~al.}(2016)\citenamefont
  {{Abbott}}, \citenamefont {{Abbott}}, \citenamefont {{Abbott}}, \citenamefont
  {{Abernathy}}, \citenamefont {{Acernese}}, \citenamefont {{Ackley}},
  \citenamefont {{Adams}}, \citenamefont {{Adams}}, \citenamefont {{Addesso}},
  \citenamefont {{Adhikari}},\ and\ \citenamefont
  {et~al.}}]{2016PhRvX...6d1015A}%
  \BibitemOpen
  \bibfield  {author} {\bibinfo {author} {\bibfnamefont {B.~P.}\ \bibnamefont
  {{Abbott}}}, \bibinfo {author} {\bibfnamefont {R.}~\bibnamefont {{Abbott}}},
  \bibinfo {author} {\bibfnamefont {T.~D.}\ \bibnamefont {{Abbott}}}, \bibinfo
  {author} {\bibfnamefont {M.~R.}\ \bibnamefont {{Abernathy}}}, \bibinfo
  {author} {\bibfnamefont {F.}~\bibnamefont {{Acernese}}}, \bibinfo {author}
  {\bibfnamefont {K.}~\bibnamefont {{Ackley}}}, \bibinfo {author}
  {\bibfnamefont {C.}~\bibnamefont {{Adams}}}, \bibinfo {author} {\bibfnamefont
  {T.}~\bibnamefont {{Adams}}}, \bibinfo {author} {\bibfnamefont
  {P.}~\bibnamefont {{Addesso}}}, \bibinfo {author} {\bibfnamefont {R.~X.}\
  \bibnamefont {{Adhikari}}}, \ and\ \bibinfo {author} {\bibnamefont
  {et~al.}},\ }\href {\doibase 10.1103/PhysRevX.6.041015} {\bibfield  {journal}
  {\bibinfo  {journal} {Physical Review X}\ }\textbf {\bibinfo {volume} {6}},\
  \bibinfo {eid} {041015} (\bibinfo {year} {2016})},\ \Eprint
  {http://arxiv.org/abs/1606.04856} {arXiv:1606.04856 [gr-qc]} \BibitemShut
  {NoStop}%
\bibitem [{\citenamefont {Abbott}\ \emph {et~al.}(2017)\citenamefont {Abbott}
  \emph {et~al.}}]{Abbott:2017vtc}%
  \BibitemOpen
  \bibfield  {author} {\bibinfo {author} {\bibfnamefont {B.}~\bibnamefont
  {Abbott}} \emph {et~al.} (\bibinfo {collaboration} {Virgo}),\ }\href
  {\doibase 10.1103/PhysRevLett.118.221101} {\bibfield  {journal} {\bibinfo
  {journal} {Phys. Rev. Lett.}\ }\textbf {\bibinfo {volume} {118}},\ \bibinfo
  {pages} {221101} (\bibinfo {year} {2017})},\ \Eprint
  {http://arxiv.org/abs/1706.01812} {arXiv:1706.01812 [gr-qc]} \BibitemShut
  {NoStop}%
\bibitem [{\citenamefont {{The LIGO Scientific Collaboration}}\ \emph
  {et~al.}(2017)\citenamefont {{The LIGO Scientific Collaboration}},
  \citenamefont {{the Virgo Collaboration}}, \citenamefont {{Abbott}},
  \citenamefont {{Abbott}}, \citenamefont {{Abbott}}, \citenamefont
  {{Acernese}}, \citenamefont {{Ackley}}, \citenamefont {{Adams}},
  \citenamefont {{Adams}}, \citenamefont {{Addesso}},\ and\ \citenamefont
  {et~al.}}]{2017arXiv170909660T}%
  \BibitemOpen
  \bibfield  {author} {\bibinfo {author} {\bibnamefont {{The LIGO Scientific
  Collaboration}}}, \bibinfo {author} {\bibnamefont {{the Virgo
  Collaboration}}}, \bibinfo {author} {\bibfnamefont {B.~P.}\ \bibnamefont
  {{Abbott}}}, \bibinfo {author} {\bibfnamefont {R.}~\bibnamefont {{Abbott}}},
  \bibinfo {author} {\bibfnamefont {T.~D.}\ \bibnamefont {{Abbott}}}, \bibinfo
  {author} {\bibfnamefont {F.}~\bibnamefont {{Acernese}}}, \bibinfo {author}
  {\bibfnamefont {K.}~\bibnamefont {{Ackley}}}, \bibinfo {author}
  {\bibfnamefont {C.}~\bibnamefont {{Adams}}}, \bibinfo {author} {\bibfnamefont
  {T.}~\bibnamefont {{Adams}}}, \bibinfo {author} {\bibfnamefont
  {P.}~\bibnamefont {{Addesso}}}, \ and\ \bibinfo {author} {\bibnamefont
  {et~al.}},\ }\href@noop {} {\bibfield  {journal} {\bibinfo  {journal} {ArXiv
  e-prints}\ } (\bibinfo {year} {2017})},\ \Eprint
  {http://arxiv.org/abs/1709.09660} {arXiv:1709.09660 [gr-qc]} \BibitemShut
  {NoStop}%
\bibitem [{\citenamefont {{Lamb}}(1989)}]{1989ESASP.296..215L}%
  \BibitemOpen
  \bibfield  {author} {\bibinfo {author} {\bibfnamefont {F.~K.}\ \bibnamefont
  {{Lamb}}},\ }in\ \href@noop {} {\emph {\bibinfo {booktitle} {Two Topics in
  X-Ray Astronomy, Volume 1: X Ray Binaries. Volume 2: AGN and the X Ray
  Background}}},\ \bibinfo {series} {ESA Special Publication}, Vol.\ \bibinfo
  {volume} {296},\ \bibinfo {editor} {edited by\ \bibinfo {editor}
  {\bibfnamefont {J.}~\bibnamefont {{Hunt}}}\ and\ \bibinfo {editor}
  {\bibfnamefont {B.}~\bibnamefont {{Battrick}}}}\ (\bibinfo {year}
  {1989})\BibitemShut {NoStop}%
\bibitem [{\citenamefont {{Bradshaw}}\ \emph {et~al.}(1999)\citenamefont
  {{Bradshaw}}, \citenamefont {{Fomalont}},\ and\ \citenamefont
  {{Geldzahler}}}]{1999ApJ...512L.121B}%
  \BibitemOpen
  \bibfield  {author} {\bibinfo {author} {\bibfnamefont {C.~F.}\ \bibnamefont
  {{Bradshaw}}}, \bibinfo {author} {\bibfnamefont {E.~B.}\ \bibnamefont
  {{Fomalont}}}, \ and\ \bibinfo {author} {\bibfnamefont {B.~J.}\ \bibnamefont
  {{Geldzahler}}},\ }\href {\doibase 10.1086/311889} {\bibfield  {journal}
  {\bibinfo  {journal} {\apjl}\ }\textbf {\bibinfo {volume} {512}},\ \bibinfo
  {pages} {L121} (\bibinfo {year} {1999})}\BibitemShut {NoStop}%
\bibitem [{\citenamefont {Wagoner}(1984)}]{1984ApJ...278..345W}%
  \BibitemOpen
  \bibfield  {author} {\bibinfo {author} {\bibfnamefont {R.~V.}\ \bibnamefont
  {Wagoner}},\ }\href@noop {} {\bibfield  {journal} {\bibinfo  {journal}
  {Astrophysical Journal}\ }\textbf {\bibinfo {volume} {278}},\ \bibinfo
  {pages} {345} (\bibinfo {year} {1984})}\BibitemShut {NoStop}%
\bibitem [{\citenamefont {{Leaci}}\ and\ \citenamefont
  {{Prix}}(2015)}]{2015PhRvD..91j2003L}%
  \BibitemOpen
  \bibfield  {author} {\bibinfo {author} {\bibfnamefont {P.}~\bibnamefont
  {{Leaci}}}\ and\ \bibinfo {author} {\bibfnamefont {R.}~\bibnamefont
  {{Prix}}},\ }\href {\doibase 10.1103/PhysRevD.91.102003} {\bibfield
  {journal} {\bibinfo  {journal} {\prd}\ }\textbf {\bibinfo {volume} {91}},\
  \bibinfo {eid} {102003} (\bibinfo {year} {2015})},\ \Eprint
  {http://arxiv.org/abs/1502.00914} {arXiv:1502.00914 [gr-qc]} \BibitemShut
  {NoStop}%
\bibitem [{\citenamefont {{Whelan}}\ \emph {et~al.}(2015)\citenamefont
  {{Whelan}}, \citenamefont {{Sundaresan}}, \citenamefont {{Zhang}},\ and\
  \citenamefont {{Peiris}}}]{2015PhRvD..91j2005W}%
  \BibitemOpen
  \bibfield  {author} {\bibinfo {author} {\bibfnamefont {J.~T.}\ \bibnamefont
  {{Whelan}}}, \bibinfo {author} {\bibfnamefont {S.}~\bibnamefont
  {{Sundaresan}}}, \bibinfo {author} {\bibfnamefont {Y.}~\bibnamefont
  {{Zhang}}}, \ and\ \bibinfo {author} {\bibfnamefont {P.}~\bibnamefont
  {{Peiris}}},\ }\href {\doibase 10.1103/PhysRevD.91.102005} {\bibfield
  {journal} {\bibinfo  {journal} {\prd}\ }\textbf {\bibinfo {volume} {91}},\
  \bibinfo {eid} {102005} (\bibinfo {year} {2015})},\ \Eprint
  {http://arxiv.org/abs/1504.05890} {arXiv:1504.05890 [gr-qc]} \BibitemShut
  {NoStop}%
\bibitem [{\citenamefont {{Messenger}}\ \emph {et~al.}(2015)\citenamefont
  {{Messenger}}, \citenamefont {{Bulten}}, \citenamefont {{Crowder}},
  \citenamefont {{Dergachev}}, \citenamefont {{Galloway}}, \citenamefont
  {{Goetz}}, \citenamefont {{Jonker}}, \citenamefont {{Lasky}}, \citenamefont
  {{Meadors}}, \citenamefont {{Melatos}}, \citenamefont {{Premachandra}},
  \citenamefont {{Riles}}, \citenamefont {{Sammut}}, \citenamefont {{Thrane}},
  \citenamefont {{Whelan}},\ and\ \citenamefont
  {{Zhang}}}]{2015PhRvD..92b3006M}%
  \BibitemOpen
  \bibfield  {author} {\bibinfo {author} {\bibfnamefont {C.}~\bibnamefont
  {{Messenger}}}, \bibinfo {author} {\bibfnamefont {H.~J.}\ \bibnamefont
  {{Bulten}}}, \bibinfo {author} {\bibfnamefont {S.~G.}\ \bibnamefont
  {{Crowder}}}, \bibinfo {author} {\bibfnamefont {V.}~\bibnamefont
  {{Dergachev}}}, \bibinfo {author} {\bibfnamefont {D.~K.}\ \bibnamefont
  {{Galloway}}}, \bibinfo {author} {\bibfnamefont {E.}~\bibnamefont {{Goetz}}},
  \bibinfo {author} {\bibfnamefont {R.~J.~G.}\ \bibnamefont {{Jonker}}},
  \bibinfo {author} {\bibfnamefont {P.~D.}\ \bibnamefont {{Lasky}}}, \bibinfo
  {author} {\bibfnamefont {G.~D.}\ \bibnamefont {{Meadors}}}, \bibinfo {author}
  {\bibfnamefont {A.}~\bibnamefont {{Melatos}}}, \bibinfo {author}
  {\bibfnamefont {S.}~\bibnamefont {{Premachandra}}}, \bibinfo {author}
  {\bibfnamefont {K.}~\bibnamefont {{Riles}}}, \bibinfo {author} {\bibfnamefont
  {L.}~\bibnamefont {{Sammut}}}, \bibinfo {author} {\bibfnamefont {E.~H.}\
  \bibnamefont {{Thrane}}}, \bibinfo {author} {\bibfnamefont {J.~T.}\
  \bibnamefont {{Whelan}}}, \ and\ \bibinfo {author} {\bibfnamefont
  {Y.}~\bibnamefont {{Zhang}}},\ }\href {\doibase 10.1103/PhysRevD.92.023006}
  {\bibfield  {journal} {\bibinfo  {journal} {\prd}\ }\textbf {\bibinfo
  {volume} {92}},\ \bibinfo {eid} {023006} (\bibinfo {year} {2015})},\ \Eprint
  {http://arxiv.org/abs/1504.05889} {arXiv:1504.05889 [gr-qc]} \BibitemShut
  {NoStop}%
\bibitem [{\citenamefont {{Aasi}}\ \emph {et~al.}(2015)\citenamefont {{Aasi}},
  \citenamefont {{Abbott}}, \citenamefont {{Abbott}}, \citenamefont {{Abbott}},
  \citenamefont {{Abernathy}}, \citenamefont {{Acernese}}, \citenamefont
  {{Ackley}}, \citenamefont {{Adams}}, \citenamefont {{Adams}}, \citenamefont
  {{Addesso}},\ and\ \citenamefont {et~al.}}]{2015PhRvD..91f2008A}%
  \BibitemOpen
  \bibfield  {author} {\bibinfo {author} {\bibfnamefont {J.}~\bibnamefont
  {{Aasi}}}, \bibinfo {author} {\bibfnamefont {B.~P.}\ \bibnamefont
  {{Abbott}}}, \bibinfo {author} {\bibfnamefont {R.}~\bibnamefont {{Abbott}}},
  \bibinfo {author} {\bibfnamefont {T.}~\bibnamefont {{Abbott}}}, \bibinfo
  {author} {\bibfnamefont {M.~R.}\ \bibnamefont {{Abernathy}}}, \bibinfo
  {author} {\bibfnamefont {F.}~\bibnamefont {{Acernese}}}, \bibinfo {author}
  {\bibfnamefont {K.}~\bibnamefont {{Ackley}}}, \bibinfo {author}
  {\bibfnamefont {C.}~\bibnamefont {{Adams}}}, \bibinfo {author} {\bibfnamefont
  {T.}~\bibnamefont {{Adams}}}, \bibinfo {author} {\bibfnamefont
  {P.}~\bibnamefont {{Addesso}}}, \ and\ \bibinfo {author} {\bibnamefont
  {et~al.}},\ }\href {\doibase 10.1103/PhysRevD.91.062008} {\bibfield
  {journal} {\bibinfo  {journal} {\prd}\ }\textbf {\bibinfo {volume} {91}},\
  \bibinfo {eid} {062008} (\bibinfo {year} {2015})},\ \Eprint
  {http://arxiv.org/abs/1412.0605} {arXiv:1412.0605 [gr-qc]} \BibitemShut
  {NoStop}%
\bibitem [{\citenamefont {{Andersson}}\ and\ \citenamefont
  {{Kokkotas}}(2001)}]{2001IJMPD..10..381A}%
  \BibitemOpen
  \bibfield  {author} {\bibinfo {author} {\bibfnamefont {N.}~\bibnamefont
  {{Andersson}}}\ and\ \bibinfo {author} {\bibfnamefont {K.~D.}\ \bibnamefont
  {{Kokkotas}}},\ }\href {\doibase 10.1142/S0218271801001062} {\bibfield
  {journal} {\bibinfo  {journal} {International Journal of Modern Physics D}\
  }\textbf {\bibinfo {volume} {10}},\ \bibinfo {pages} {381} (\bibinfo {year}
  {2001})},\ \Eprint {http://arxiv.org/abs/gr-qc/0010102} {gr-qc/0010102}
  \BibitemShut {NoStop}%
\bibitem [{\citenamefont {{Lindblom}}\ \emph {et~al.}(2000)\citenamefont
  {{Lindblom}}, \citenamefont {{Owen}},\ and\ \citenamefont
  {{Ushomirsky}}}]{2000PhRvD..62h4030L}%
  \BibitemOpen
  \bibfield  {author} {\bibinfo {author} {\bibfnamefont {L.}~\bibnamefont
  {{Lindblom}}}, \bibinfo {author} {\bibfnamefont {B.~J.}\ \bibnamefont
  {{Owen}}}, \ and\ \bibinfo {author} {\bibfnamefont {G.}~\bibnamefont
  {{Ushomirsky}}},\ }\href {\doibase 10.1103/PhysRevD.62.084030} {\bibfield
  {journal} {\bibinfo  {journal} {\prd}\ }\textbf {\bibinfo {volume} {62}},\
  \bibinfo {eid} {084030} (\bibinfo {year} {2000})},\ \Eprint
  {http://arxiv.org/abs/astro-ph/0006242} {astro-ph/0006242} \BibitemShut
  {NoStop}%
\bibitem [{\citenamefont {{Lewin}}\ \emph {et~al.}(1995)\citenamefont
  {{Lewin}}, \citenamefont {{van Paradijs}},\ and\ \citenamefont {{van den
  Heuvel}}}]{1995xrbi.nasa.....L}%
  \BibitemOpen
  \bibfield  {author} {\bibinfo {author} {\bibfnamefont {W.~H.~G.}\
  \bibnamefont {{Lewin}}}, \bibinfo {author} {\bibfnamefont {J.}~\bibnamefont
  {{van Paradijs}}}, \ and\ \bibinfo {author} {\bibfnamefont {E.~P.~J.}\
  \bibnamefont {{van den Heuvel}}},\ }\href@noop {} {\bibfield  {journal}
  {\bibinfo  {journal} {X-ray Binaries}\ } (\bibinfo {year}
  {1995})}\BibitemShut {NoStop}%
\bibitem [{\citenamefont {{McNamara}}\ \emph {et~al.}(2005)\citenamefont
  {{McNamara}}, \citenamefont {{Norwood}}, \citenamefont {{Harrison}},
  \citenamefont {{Holtzman}}, \citenamefont {{Dukes}},\ and\ \citenamefont
  {{Barker}}}]{2005ApJ...623.1070M}%
  \BibitemOpen
  \bibfield  {author} {\bibinfo {author} {\bibfnamefont {B.~J.}\ \bibnamefont
  {{McNamara}}}, \bibinfo {author} {\bibfnamefont {J.}~\bibnamefont
  {{Norwood}}}, \bibinfo {author} {\bibfnamefont {T.~E.}\ \bibnamefont
  {{Harrison}}}, \bibinfo {author} {\bibfnamefont {J.}~\bibnamefont
  {{Holtzman}}}, \bibinfo {author} {\bibfnamefont {R.}~\bibnamefont {{Dukes}}},
  \ and\ \bibinfo {author} {\bibfnamefont {T.}~\bibnamefont {{Barker}}},\
  }\href {\doibase 10.1086/428640} {\bibfield  {journal} {\bibinfo  {journal}
  {\apj}\ }\textbf {\bibinfo {volume} {623}},\ \bibinfo {pages} {1070}
  (\bibinfo {year} {2005})}\BibitemShut {NoStop}%
\bibitem [{\citenamefont {{Mukherjee}}\ \emph {et~al.}(2015)\citenamefont
  {{Mukherjee}}, \citenamefont {{Bult}}, \citenamefont {{van der Klis}},\ and\
  \citenamefont {{Bhattacharya}}}]{2015MNRAS.452.3994M}%
  \BibitemOpen
  \bibfield  {author} {\bibinfo {author} {\bibfnamefont {D.}~\bibnamefont
  {{Mukherjee}}}, \bibinfo {author} {\bibfnamefont {P.}~\bibnamefont {{Bult}}},
  \bibinfo {author} {\bibfnamefont {M.}~\bibnamefont {{van der Klis}}}, \ and\
  \bibinfo {author} {\bibfnamefont {D.}~\bibnamefont {{Bhattacharya}}},\ }\href
  {\doibase 10.1093/mnras/stv1542} {\bibfield  {journal} {\bibinfo  {journal}
  {\mnras}\ }\textbf {\bibinfo {volume} {452}},\ \bibinfo {pages} {3994}
  (\bibinfo {year} {2015})},\ \Eprint {http://arxiv.org/abs/1507.02138}
  {arXiv:1507.02138 [astro-ph.HE]} \BibitemShut {NoStop}%
\bibitem [{\citenamefont {{Papitto}}\ \emph {et~al.}(2013)\citenamefont
  {{Papitto}}, \citenamefont {{D'A{\`i}}}, \citenamefont {{Di Salvo}},
  \citenamefont {{Egron}}, \citenamefont {{Bozzo}}, \citenamefont {{Burderi}},
  \citenamefont {{Iaria}}, \citenamefont {{Riggio}},\ and\ \citenamefont
  {{Menna}}}]{2013MNRAS.429.3411P}%
  \BibitemOpen
  \bibfield  {author} {\bibinfo {author} {\bibfnamefont {A.}~\bibnamefont
  {{Papitto}}}, \bibinfo {author} {\bibfnamefont {A.}~\bibnamefont
  {{D'A{\`i}}}}, \bibinfo {author} {\bibfnamefont {T.}~\bibnamefont {{Di
  Salvo}}}, \bibinfo {author} {\bibfnamefont {E.}~\bibnamefont {{Egron}}},
  \bibinfo {author} {\bibfnamefont {E.}~\bibnamefont {{Bozzo}}}, \bibinfo
  {author} {\bibfnamefont {L.}~\bibnamefont {{Burderi}}}, \bibinfo {author}
  {\bibfnamefont {R.}~\bibnamefont {{Iaria}}}, \bibinfo {author} {\bibfnamefont
  {A.}~\bibnamefont {{Riggio}}}, \ and\ \bibinfo {author} {\bibfnamefont
  {M.~T.}\ \bibnamefont {{Menna}}},\ }\href {\doibase 10.1093/mnras/sts605}
  {\bibfield  {journal} {\bibinfo  {journal} {\mnras}\ }\textbf {\bibinfo
  {volume} {429}},\ \bibinfo {pages} {3411} (\bibinfo {year} {2013})},\ \Eprint
  {http://arxiv.org/abs/1212.2532} {arXiv:1212.2532 [astro-ph.HE]} \BibitemShut
  {NoStop}%
\bibitem [{\citenamefont {Illarionov}\ and\ \citenamefont
  {Sunyaev}(1975)}]{Illarionov:1975ei}%
  \BibitemOpen
  \bibfield  {author} {\bibinfo {author} {\bibfnamefont {A.~F.}\ \bibnamefont
  {Illarionov}}\ and\ \bibinfo {author} {\bibfnamefont {R.~A.}\ \bibnamefont
  {Sunyaev}},\ }\href@noop {} {\bibfield  {journal} {\bibinfo  {journal}
  {Astron. Astrophys.}\ }\textbf {\bibinfo {volume} {39}},\ \bibinfo {pages}
  {185} (\bibinfo {year} {1975})}\BibitemShut {NoStop}%
\bibitem [{\citenamefont {{Cook}}\ \emph {et~al.}(1994)\citenamefont {{Cook}},
  \citenamefont {{Shapiro}},\ and\ \citenamefont
  {{Teukolsky}}}]{1994ApJ...424..823C}%
  \BibitemOpen
  \bibfield  {author} {\bibinfo {author} {\bibfnamefont {G.~B.}\ \bibnamefont
  {{Cook}}}, \bibinfo {author} {\bibfnamefont {S.~L.}\ \bibnamefont
  {{Shapiro}}}, \ and\ \bibinfo {author} {\bibfnamefont {S.~A.}\ \bibnamefont
  {{Teukolsky}}},\ }\href {\doibase 10.1086/173934} {\bibfield  {journal}
  {\bibinfo  {journal} {\apj}\ }\textbf {\bibinfo {volume} {424}},\ \bibinfo
  {pages} {823} (\bibinfo {year} {1994})}\BibitemShut {NoStop}%
\end{thebibliography}%

\end{document}